\documentclass[lettersize,journal]{IEEEtran}

\usepackage{graphicx}
\usepackage{url}
\usepackage{color}
\usepackage{booktabs}
\usepackage{fontawesome}
\usepackage{amsmath}
\usepackage{tabularx}  
\usepackage{float}
\usepackage{subfig}  
\usepackage{rotating}    
\usepackage{makecell}    
\usepackage{multicol} 
\usepackage{multirow} 
\usepackage[ruled]{algorithm2e}
\usepackage{xcolor}
\usepackage{pifont} 
\usepackage{colortbl} 
\usepackage{algpseudocode}
\usepackage{tikz}
\usepackage{filecontents}
\usepackage{diagbox}
\usepackage{etoolbox} 

\newcommand\fig{Fig.}
\newcommand\tab{Tab.}
\newcommand\alg{Algorithm.}
\newcommand\eqt{Eq.}

\newcommand\Sec{Section.}

\newcommand{\udm}[1]{\textcolor{black}{#1}}

\begin{document}
\title{BlockFlex: A Hybrid Resilient Routing Method based on Robust Virtual Overlay for Mega-constellation Networks}

\author{
	\IEEEauthorblockN{
		Wang~Xiangtong
		}




		
}
\markboth{Journal of \LaTeX\ Class Files,~Vol.~14, No.~8, August~2030}%
{Shell \MakeLowercase{\textit{et al.}}: A Sample Article Using IEEEtran.cls for IEEE Journals}

\maketitle

\begin{abstract}

Mega-constellation networks, comprising thousands of interconnected low-Earth-orbit (LEO) satellites, represent a transformative leap in global Internet connectivity.
However, their operational promise is constrained by complex dynamics arising from persistent satellite--ground topology changes and intermittent inter-satellite link (ISL) failures.
These dynamics simultaneously trigger global control-plane overhead and induce sparse connectivity in the network, collectively degrading both the efficiency and resilience of routing.
To address these challenges, this paper proposes BlockFLEX, a hybrid routing architecture built upon a robust virtual overlay for LEO mega-constellation networks.
BlockFLEX constructs a robust virtual overlay by clustering satellites into anonymous \emph{blocks}, which masks underlying network dynamics and provides a stable topology view for the routing layer.
The architecture further employs a two-tier hybrid routing strategy: convergence-free geographic forwarding operates \emph{between} blocks, while convergence-isolated routing runs \emph{within} each block, thereby localizing control-traffic propagation.
Beyond this core routing foundation, BlockFLEX incorporates complementary mechanisms to enhance scalability, resilience, and efficiency.
Experimental evaluations on current operational LEO mega-constellation networks demonstrate that, under scenarios with up to $30\%$ random ISL failures, BlockFLEX substantially outperforms state-of-the-art schemes in both routing resilience and efficiency.

\end{abstract}


\section{Introduction}

Low Earth Orbit (LEO) mega-constellation networks (MCNs) have recently emerged as a pivotal infrastructure to complement terrestrial networks, aiming to deliver ubiquitous, high-bandwidth, and low-latency Internet connectivity to unserved regions, maritime, and aviation sectors. Unlike Geostationary Earth Orbit (GEO) systems \cite{chinasat2025}, LEO MCNs leverage dense satellite deployments and Inter-Satellite Links (ISLs) to achieve global coverage. Recent commercial successes~\cite{starlink,oneweb,kuiper,telesat} and new satellite architectures \cite{karkadakattil2025ai} have further validated the feasibility of large-scale MCNs, fundamentally reshaping the global connectivity paradigm.

Despite their significant promise, MCNs introduce unique challenges for routing design due to their complex dynamics.
First, LEO satellites travel at roughly $7.6\,\mathrm{km/s}$ relative to the ground~\cite{bhattacherjee2019network}, creating satellite--ground topology dynamics that change far more rapidly than in terrestrial or geostationary networks. This rapid motion leads to frequent handovers of Ground-to-Satellite Links (GSLs), further compounding the routing challenge.
Second, the ISLs that support long-distance transmission services face several sources of disruption. Natural events such as sun outages and orbital maneuvers, along with satellite motion, can cause temporary link unavailability~\cite{eguri2024survey}.
More critically, ISLs are also vulnerable to deliberate disruptions, including Denial-of-Service (DoS) attacks, whose resulting link unavailability can be treated as ISL failures~\cite{giuliari2021icarus}.

Previous research addressing network dynamics has primarily focused on modifying routing convergence mechanisms.
For example, existing studies rely on orbit prediction~\cite{pan2019opspf,fischer2012predictable,rfc9717,ruan2022lightweight} or geographic routing~\cite{navas1997geocast,henderson2000distributed,li2024stable,tsunoda2006geographical} to eliminate the Link-State-Advertisement (LSA) overhead caused by satellite--ground topology changes.
Alternatively, some studies adopt a centralized control plane~\cite{Peng2024FastTSEF,lai2023achieving,de2025skylink} to cope with the LSA flooding triggered by ISL failures.
By taking a global, topology-aware perspective to counteract dynamics, these methods maintain routing consistency, thereby ensuring network reachability and resilience.
However, their topology-awareness mechanisms introduce substantial overhead, such as large-scale ephemeris synchronization or coordination within a centralized control plane.
While geographic routing~\cite{navas1997geocast,henderson2000distributed,li2024stable,tsunoda2006geographical} offers clear efficiency advantages by avoiding control overhead, the sparse connectivity caused by ISL failures can lead to unreachable paths and, consequently, degrade routing resilience.
As a result, existing solutions struggle to simultaneously achieve high network resilience and efficient routing performance.
This challenge is further amplified by emerging networking scenarios characterized by massive long-distance traffic and complex mobility patterns, such as international aviation communication or remote-sensing satellite data transmission.
These limitations underscore the need for a new routing architecture capable of natively balancing efficiency and resilience within highly dynamic MCNs.

The core limitation of existing approaches lies in the global propagation of localized disturbances, combined with the sparse connectivity induced by ISL failures, which collectively undermine routing efficiency and resilience.
To bridge this gap, this paper proposes BlockFlex, a robust and hybrid routing architecture that flexibly organizes satellites into autonomous domains (blocks) to achieve routing efficiency and resilience in highly dynamic mega-constellation networks.
The design philosophy of BlockFlex centers on \emph{dynamics localization} and \emph{connectivity enhancement}: by partitioning the constellation into autonomous domains, BlockFlex decouples the inherent instability of the physical underlay from the routing logic, while simultaneously reinforcing inter-domain connectivity to create a more stable virtual overlay for routing.
To realize this philosophy, BlockFlex incorporates two synergistic components:

\begin{itemize}
	\item \textbf{A robust virtual overlay network}. This component masks network dynamics from the routing layer and enhances connectivity, providing a stable and robust topology abstraction for the upper-layer routing scheme.
	\item \textbf{A hybrid routing scheme}. Building atop the virtual overlay, BlockFlex applies different routing strategies \emph{between} and \emph{within} blocks, and seamlessly integrates them to complete end-to-end delivery, jointly ensuring both routing efficiency and resilience in a fully distributed manner.
\end{itemize}

In summary, the main contributions of this work are as follows:

\noindent\ding{117}~A robust virtual overlay network is designed to effectively mask the impact of intermittent ISL failures on global routing. By confining topological instability within autonomous blocks, it maintains high network robustness even under severe and unpredictable failure conditions.

\noindent\ding{117}~A hybrid routing scheme is proposed that balances routing efficiency and resilience by applying different packet forwarding strategies between and within blocks, and seamlessly integrating them to complete end-to-end delivery in a fully distributed manner.

\noindent\ding{117}\udm{~BlockFlex is presented as a system-level architecture that integrates virtual overlay and hybrid routing into a cohesive design. While domain-based partitioning, geographic forwarding, and hierarchical routing have been explored individually, BlockFlex is, to the best of our knowledge, the first to synergistically combine them for high resilience and efficiency under severe ISL failures.}

\noindent\ding{117}~BlockFlex is evaluated on operational mega-constellations using an open-source testbed. Under scenarios with up to $30\%$ unpredictable ISL failures, BlockFlex achieves a $2\times$ improvement in routing reachability compared to state-of-the-art schemes while maintaining near-$100\%$ delivery. The control-message and FIB-update overhead remains below $0.2\%$ of that required by convergent protocols such as OSPF. Additionally, BlockFlex reduces routing-computation time by $\geq 36\%$ and latency jitter by $\geq 61\%$.

\section{Background and Related Work}

\subsection{Mega-constellation Network Architecture.}

\begin{figure}[t!]
    \centering
    \begin{minipage}{0.48\textwidth}
        \centering
        \includegraphics[width=\linewidth]{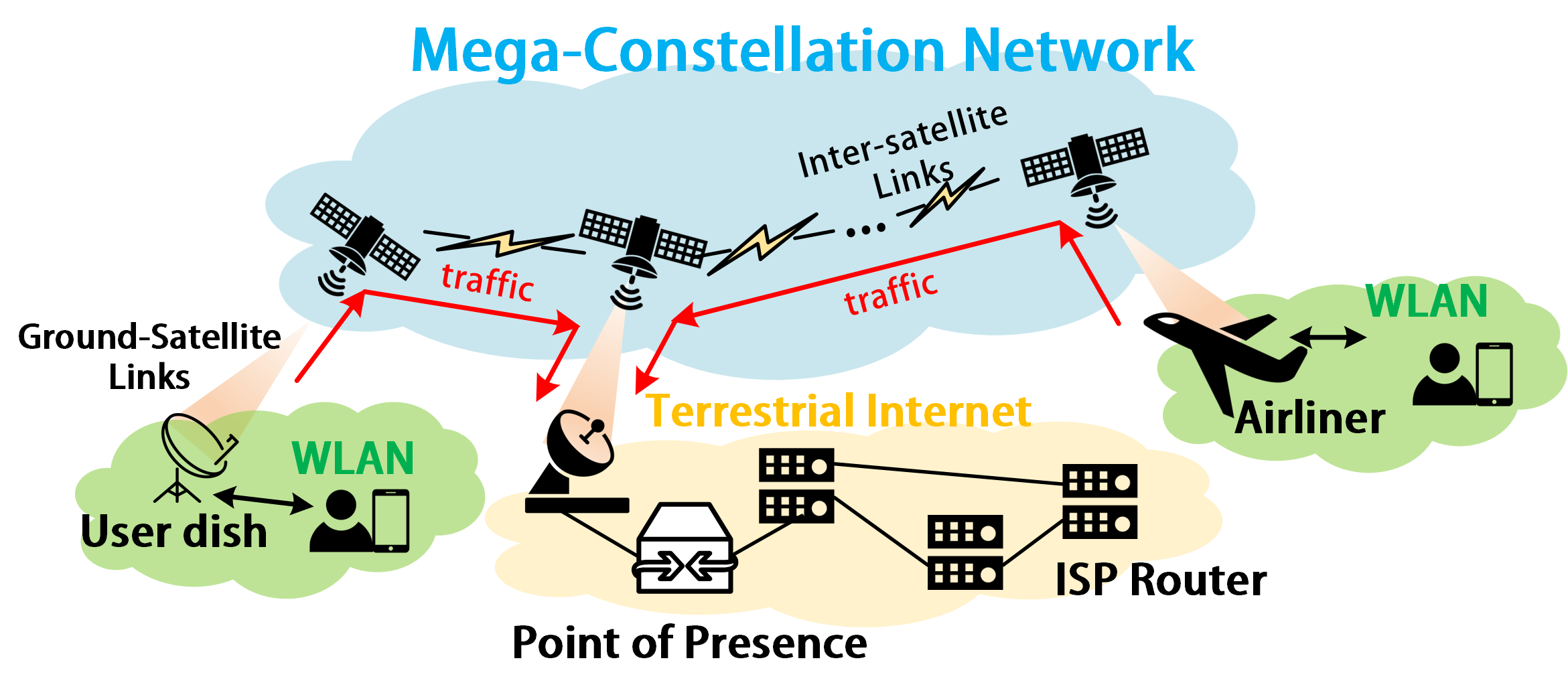}
        \caption{Today's Mega-constellation network architecture.}
        \vspace{-1em}
        \label{fig:arch}
    \end{minipage}%
\end{figure}

\fig\ref{fig:arch} illustrates the architecture of mega-constellation networks, comprising space, ground, and user segments. 
The space segment consisting of thousands of satellites, enabling global connectivity via laser ISLs. 
The ground segment consisting of an operational terrestrial network with Ground Stations (GS), and 
the user segment consisting of user dishes, terminals or in‑flight  WLAN \cite{StarlinkTechOverview}.
In operation, a user transmits data via a dish to a source satellite over a GSL. The traffic is then relayed across the constellation through multiple ISLs until it reaches the destination satellite. This satellite downlinks the data to a GS, which forwards it to a connected Point of Presence for final internet access.
Both GSes and user dishes function as the edge nodes of the network. They are responsible for protocol conversion and adapting terrestrial data formats for transmission over the satellite links, thereby bridging the space and ground segments.

\subsection{Related work}
\label{sec:related}

\subsubsection{MCN Routing}

Routing in MCNs must cope with highly frequent GSL handovers and random ISL disruptions. Traditional terrestrial protocols (e.g., Open Shortest Path First (OSPF)\cite{rfc5340}, Border Gateway Protocol (BGP)\cite{rfc4271}, Ad Hoc On-Demand Distance Vector Routing (AODV)\cite{perkins2003rfc3561}) achieve topology consistency via LSA. However, the resulting global convergence incurs prohibitive control overhead in large-scale MCNs due to frequent link-state updates \cite{tanveer2023making}.

To mitigate this, one approach leverages the predictable mesh‑like inter‑satellite topology for satellite-to-satellite routing through structured addressing \cite{ekici2000datagram,chen2019distributed,Kedrowitsch2024ResilientRF}. While efficient for the space segment, end‑to‑end delivery still requires identifying which satellite currently covers a given destination. Ephemeris prediction methods \cite{pan2019opspf,fischer2012predictable,rfc9717} have been proposed to bridge this gap without massive signaling. Nevertheless, these methods require continuous ephemeris synchronization at network endpoints and offer limited support for highly mobile users.

Alternatively, centralized control‑plane approaches (e.g., OpenSAN \cite{bao2014opensan}, SDSN \cite{xie2019sdsn}) collect local states at a central controller to compute global routes. This eliminates inter-satellite control exchange and enables centralized fault‑tolerance \cite{feng2022dr,lai2023achieving}. However, the signaling between the data and control planes still scales with network size, leading to substantial overhead as constellations expand. Moreover, the fixed placement of controllers constrains flexibility in MCN for mobile terminals. Recent studies \cite{navas1997geocast,ekici2000datagram,lai2021orbitcast,li2024stable} have explored stateless, convergence-free geographic routing to improve scalability. Yet, as evaluated in \Sec\ref{exp:eff} and \Sec\ref{exp:res}, these solutions often fail to balance efficiency and resilience simultaneously under the inherent dynamics of MCNs.

\subsubsection{Masking the Dynamics for MCNs Routing}
Volatility, including frequent handovers and intermittent ISL failures, presents a fundamental challenge. To mask this, virtual topology or virtual node abstractions \cite{korccak2009virtual,chen2019topology,lu2013virtual} replace global convergence with localized container switching. While this simplifies the problem to a near-static level, the latency and bandwidth overhead of mapping remains a challenge.
Another line of work treats random ISL failures as burst traffic \cite{lai2023achieving} or introduces Dynamic Discrete Topology \cite{li2024dynamic} as a virtual overlay. However, these methods often rely on ground-based centralized controllers to aggregate failure information, limiting their adaptability in unpredictable environments. BlockFlex addresses these limitations through its DABNet overlay, which localizes and mitigates the impact of dynamics on routing in a distributed manner to enhance robustness.


\subsubsection{Domain-based Satellite Routing}
Partitioning large‑scale networks into domains effectively reduces global routing overhead \cite{haas1997new}. This concept has been applied in wireless mobile networks via clustering \cite{krishna1997cluster,dang2010clustering} and extended to satellite routing \cite{li2023leo,liu2021reliable,du2025multi,liu2025leo}.
For example, Du et al.~\cite{du2025multi} proposed an orbit-based hierarchical routing scheme that partitions the network via a predefined orbital hierarchy; the segmentation is updated only when newly deployed satellites join the constellation. While this approach reduces global routing overhead, its static division structure struggles to adapt to frequent ISL failures and dynamic topology changes.
Liu et al.~\cite{liu2025leo} adopted a dynamic partitioning strategy in which each domain maintains two edge-disjoint paths. Both intra-domain and inter-domain routing rely on LSA-based mechanisms. However, their partitioning method requires a centralized control strategy, and the inter-domain LSA mechanism still incurs global flooding overhead.
In contrast, the domains (blocks) in BlockFlex do not have fixed community. Instead, they evolve adaptively based on dynamic topology caused by ISL failures of MCN. BlockFlex also adopt a hybrid routing scheme that combines convergence-free geographic forwarding between blocks and convergence-isolated routing within blocks, which not only reduces routing and failure‑response overhead, but more importantly, sustains high‑degree connectivity between blocks, thereby establishing a robust foundation for routing operation.

\section{System Overview}
\label{sec:sys}

\begin{figure}[t!]
    \centering
        \includegraphics[width=0.97\linewidth]{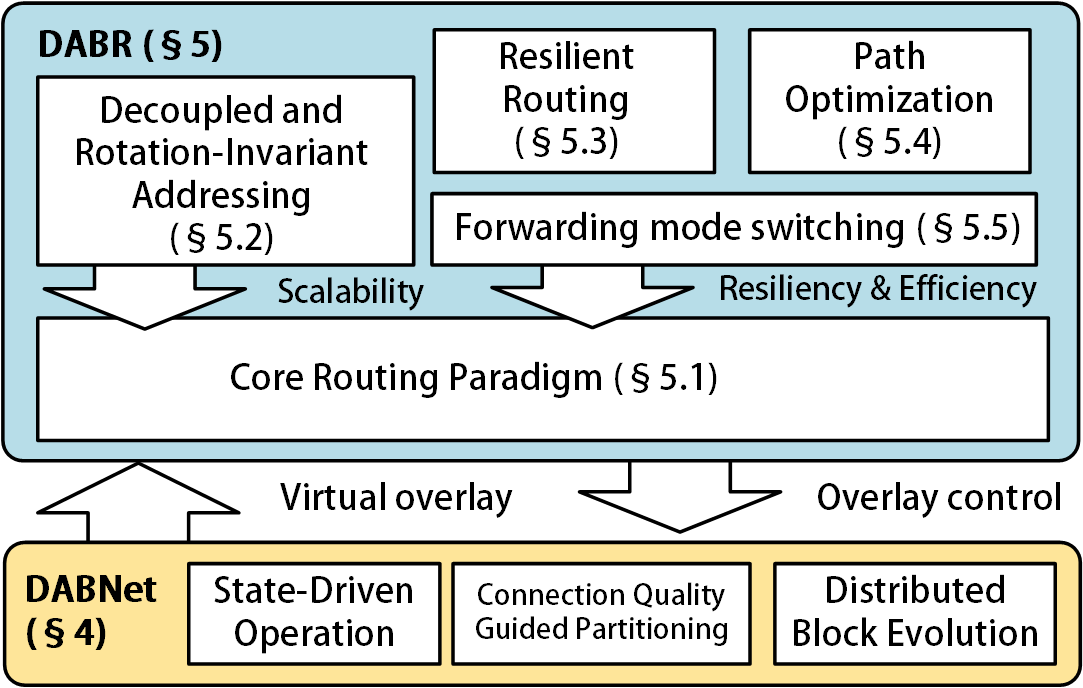}
        \caption{\udm{An overview of BlockFlex.}}
    \vspace{-1em}
        \label{fig:wkfl}
\end{figure}

This section presents a high-level overview of BlockFlex's design, as illustrated in
\fig\ref{fig:wkfl}. The key components of BlockFlex are summarized below and elaborated in the following sections.

\noindent\textbf{(1) Dynamic adaptive block network based virtual overlay} (\Sec\ref{sec:dabnet}).
A Dynamic Adaptive Block Network (DABNet) is abstracted atop the conventional MCN physical architecture as the virtual overlay in BlockFlex. 
DABNet adopts a block-based organization that localizes routing flooding caused by ISL failures within bounded scopes and enhances connectivity density among forwarding units, thereby establishing a robust, failure-transparent virtual substrate for routing operation.

\noindent\textbf{(2) Hybrid routing scheme for DABNet} (\Sec\ref{sec:dabr}).
BlockFlex employs a hybrid routing scheme (\textbf{DABR}) as its overlay operating atop DABNet.
Built upon DABNet's robust architecture, the core routing paradigm in DABR implements a two‑tier forwarding strategy in which inter‑block routing employs convergence‑free geographic forwarding across block boundaries, while intra‑block routing operates convergence‑isolated through within each block. This separation allows each block to maintain its own localized control plane, thereby containing excessive control‑traffic overhead across the network. 

Beyond this core routing foundation, BlockFlex incorporates additional components designed to enhance scalability, resilience, and latency performance. These enhancements are detailed in \Sec\ref{sec:addressing} $\sim$ \Sec\ref{sec:dual}.

\section{Dynamic Adaptive Block Network based Virtual Overlay}
\label{sec:dabnet}



The physical underlay of MCNs is inherently highly dynamic, and ISLs may experience unpredictable intermittent disruptions due to various factors~\cite{choudhary2024inter,saeed2021point,kaymak2018survey}.
To mitigate the direct impact of such topology fluctuations on the routing process, the core design principle of BlockFlex is to decouple the underlay topology from the routing logic.
Following this principle, BlockFlex introduces the DABNet as a virtual overlay atop the physical MCN underlay. DABNet adaptively maintains a relatively stable topology view and sufficient connectivity under network fluctuations, thereby providing a robust substrate for routing. It comprises the following entities:

\begin{itemize}
    \item \textbf{Autonomous blocks}: Groups of neighboring satellites that are tightly interconnected via ISLs and managed as a single logical unit. A block absorbs internal dynamics (e.g., ISL failures among its members) locally, shielding the rest of the network from their impact.
    \item \textbf{Vagrant satellites}: Satellites that currently belong to no block. They serve as candidates for future block expansion.
    \item \textbf{Forwarding units (FUs)}: The primary routing nodes in the DABNet overlay. An FU can be either a block or a vagrant satellite---from the routing perspective, they are treated uniformly.
    \item \textbf{Inter-unit links (IULs)}: Logical links $\mathcal{E}_{IUL}$ that interconnect FUs, forming the overlay topology over which inter-block routing operates.
\end{itemize}

\begin{figure}[t!]
    \centering
    \begin{minipage}{0.5\textwidth}
        \centering
        \includegraphics[width=\linewidth]{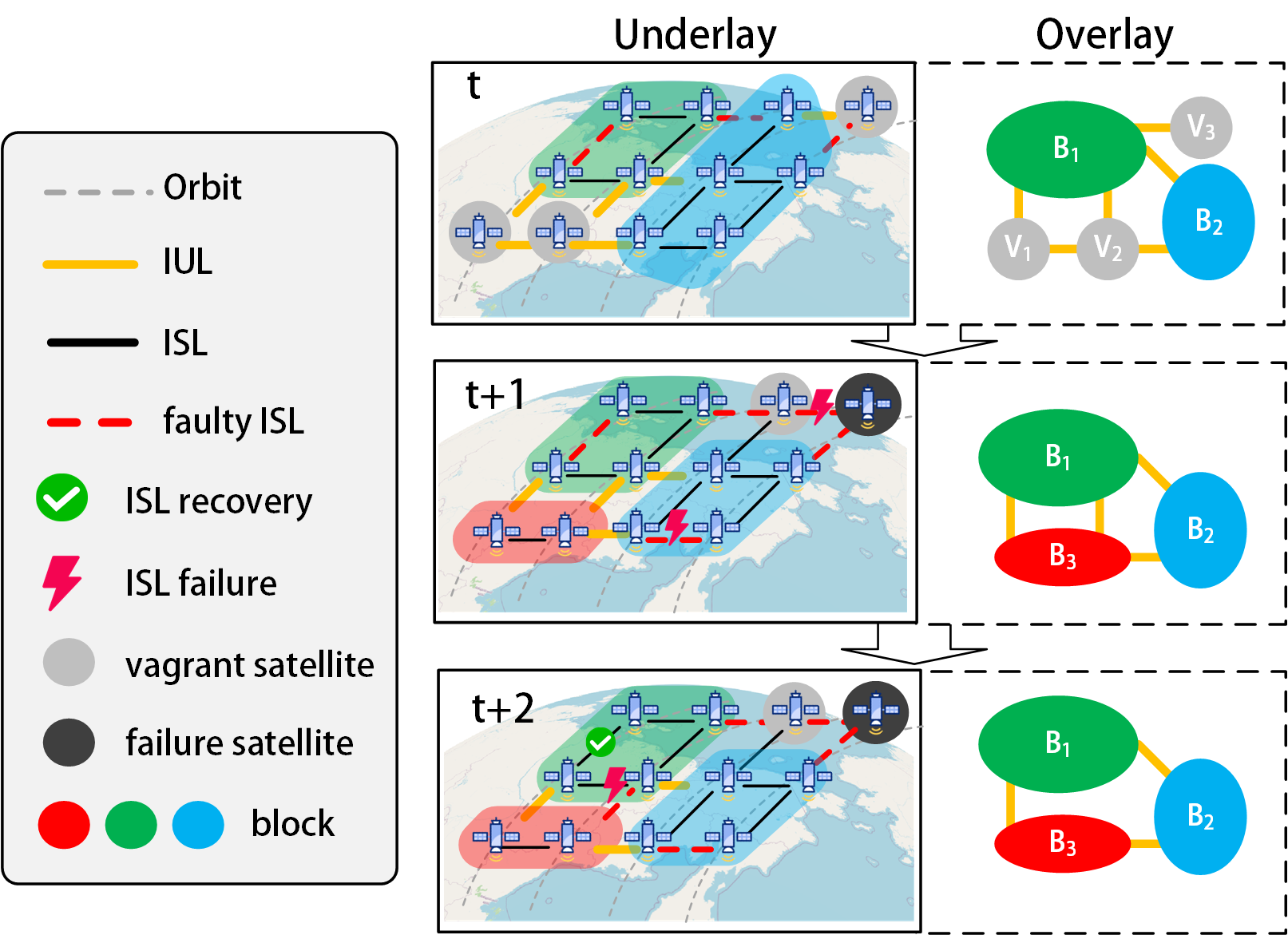}
        \caption{\udm{The DABNet architecture.}}
        \label{fig:dabnet}
    \end{minipage}%
\end{figure}

\fig\ref{fig:dabnet} illustrates the architectural composition of DABNet and its dynamic self-adaptive process.
At time $t$, the satellite underlay is partitioned into autonomous blocks (colored) and vagrant satellites (grey).
At time $t+1$, $v_1$ and $v_2$ re-aggregate to form a new autonomous block. Meanwhile, $v_3$ loses its last remaining link and becomes a faulty satellite (black), thereby leaving the DABNet topology. An ISL failure occurs within block $\mathcal{B}_2$, but since it does not disrupt $\mathcal{B}_2$'s internal connectivity, $\mathcal{B}_2$ remains unchanged from DABNet's perspective.
At time $t+2$, the failed link inside $\mathcal{B}_1$ recovers, while a link between $\mathcal{B}_3$ and $\mathcal{B}_1$ fails---neither event alters the DABNet topology.
This adaptive process demonstrates that intra-block ISL failures and recoveries are transparent to DABNet's global view, effectively confining the propagation of fault perturbations. More importantly, compared to individual satellites, autonomous blocks are typically connected by denser inter-block links, further enhancing the overlay's robustness under link fluctuations.

\subsection{State-Driven Block Evolution}

When ISL failures occur, they may disrupt the internal and external connectivity of existing blocks, degrading the robustness of DABNet. To counter this, blocks adaptively evolve in response to connectivity perturbations, thereby preserving network stability. Unlike static partitioning methods~\cite{krishna1997cluster,dang2010clustering,li2023leo,du2025multi}, each block in DABNet is governed by a decentralized Finite State Machine (FSM) that drives its evolution.

\begin{figure}[t!] 
    \centering  
      
            \begin{minipage}{0.45\textwidth} 
                \includegraphics[width=1\linewidth]{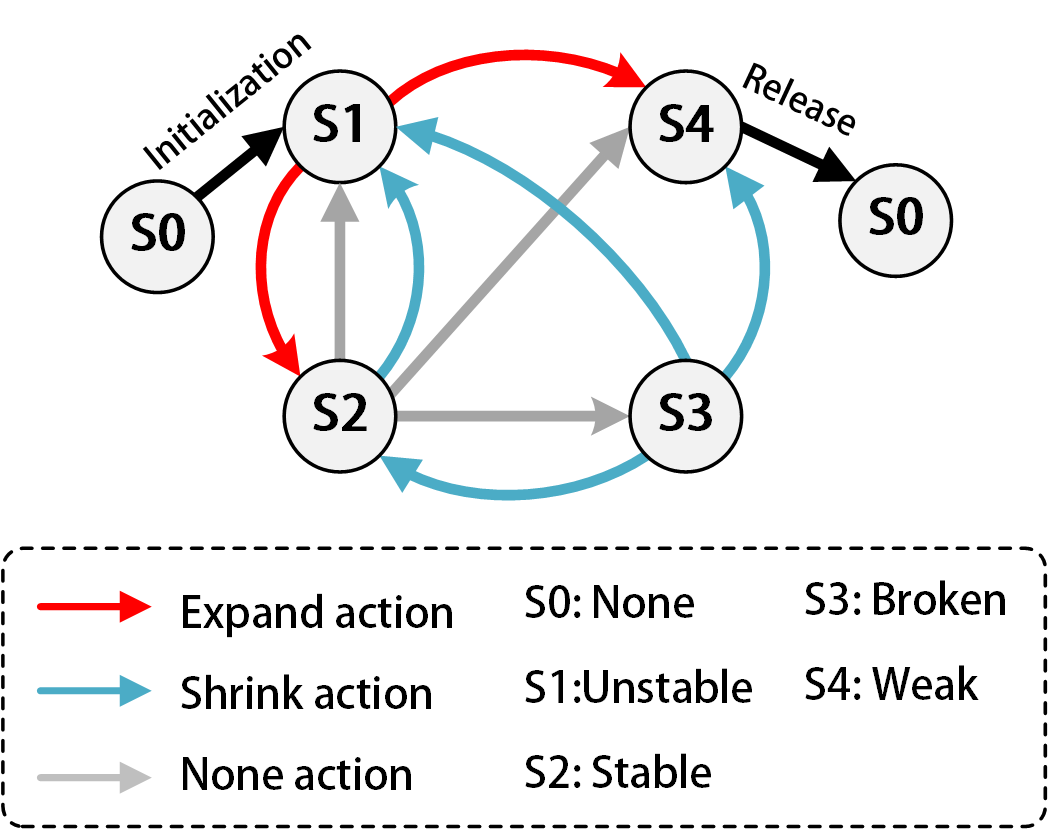} 
            \end{minipage}

    \caption{\udm{The finite state machine model for blocks in DABNet.}}
    \label{fig:states}
\end{figure}

\fig\ref{fig:states} illustrates the state definitions and state-transition diagram for autonomous blocks. This FSM model, triggered by ISL failure events, enables the distributed self-adaptive evolution of blocks in mega-constellation networks. It defines five states:
\begin{itemize}
    \item \texttt{NONE}: The block has not yet been formed or has been dissolved.
    \item \texttt{UNSTABLE}: The block size is below the minimum threshold but mergeable vagrant satellites are available, or the block diameter exceeds the upper limit but can be corrected via shrinking.
    \item \texttt{STABLE}: The block has reached its size limit or no mergeable vagrant satellites are nearby, so no further evolution is possible.
    \item \texttt{BROKEN}: The satellites in the block have been disconnected by ISL failures, rendering the block internally non-connected.
    \item \texttt{WEAK}: The block size is below the threshold and no further expansion is possible.
\end{itemize}
Based on these states, the key lifecycle transitions of an autonomous block are completed through four main actions:
\begin{itemize}
\item \textbf{Initialization}: Initially, all satellite nodes are vagrant; each satellite may become a block center through initialization with some probability. If an initialized block is unstable, it can enter a stable state through expansion.
\item \textbf{Expand}: At each time step, a block checks whether there are adjacent vagrant satellites and decides how to expand according to the evolution algorithm.
\item \textbf{Shrink}: When satellite or ISL failures occur in the network, a block may perform a shrinking operation.
\item \textbf{Release}: If a block's size falls below the limit and it cannot expand further, all member satellites are released and return to the vagrant state.
\end{itemize}

\begin{algorithm}[t!]
    \SetKwFunction{EVO}{BlockEvo}
    \SetKwProg{Fn}{def}{\string :}{}

    \caption{State-Driven Autonomous Block Evolution Framework}
    \label{alg:evo}

    \LinesNumbered 
    \KwIn{Autonomous block $\mathcal{B}^t$ at time $t$}
    \KwOut{Autonomous block $\mathcal{B}^{t+1}$ at time $t+1$}

    \Fn{\EVO{ $\mathcal{B}$: $\mathcal{B}^t$} $\to$ $\mathcal{B}^{t+1}$}{
        $\mathcal{B}^\star \leftarrow \mathcal{B}^t$;\\
        \While{\textup{\textbf{true}} }{

            \uIf{\texttt{\textup{STATE}}($\mathcal{B}^\star$) = \texttt{\textup{BROKEN}} }{
                \tcp{shrink to preserve maximum connected component}
                $\mathcal{B}^\prime \leftarrow \mathcal{B}^\star$ - \textup{\texttt{MaxSub($\mathcal{B}^\star$)}};\\
                \lFor{$s \in \mathcal{B}^\prime$ }{
                    $s$.\texttt{\textup{STATE}} $\leftarrow$ \texttt{\textup{VAGRANT}}
                }
                $\mathcal{B}^\star \leftarrow $ \textup{\texttt{MaxSub($\mathcal{B}^\star$)}};\\
            }

            \uElseIf{\texttt{\textup{STATE($\mathcal{B}^\star$)}} = \texttt{\textup{WEAK}} }{
                \tcp{relase block and all satellites become vagrant}
                \lFor{$s \in \mathcal{B}^\star$ }{$s$.\texttt{\textup{STATE}} $\leftarrow$ \texttt{\textup{VAGRANT}}}
                $\mathcal{B}^\star \leftarrow \emptyset$;~\textup{\textbf{break}};
            }

            \uElseIf{\texttt{\textup{STATE($\mathcal{B}^\star$)}} = \textup{\texttt{STABLE}}  }{
                \textup{\textbf{break}}; 
            }\ElseIf{\texttt{\textup{STATE}}($\mathcal{B}^\star$) = \texttt{\textup{UNSTABLE}}}{
                \tcp{block evolve via CQSBE algorithm}
                $\mathcal{B}^\star \leftarrow $\textup{\texttt{CQSBE($\mathcal{B}^\star$)}};
            }

        }
        $\mathcal{B}^{t+1}\leftarrow \mathcal{B}^\star $;
        \Return{$\mathcal{B}^{t+1}$}
    }
    \textbf{end}\\
\end{algorithm}

Based on the state and action definitions above, \alg\ref{alg:evo} illustrates the evolution framework of an autonomous block, where the function \textup{\texttt{BlockEvo()}} governs the entire evolution process.
When a block enters the \texttt{BROKEN} state (lines 4--8), the algorithm preserves its maximum connected component to reconstruct a new block, while the remaining satellites transition to the vagrant state.
If the block becomes \texttt{WEAK} (lines 9--12), all associated satellites are released and become available for merging.
Line 13 marks the transition to the \texttt{STABLE} state and terminates the iteration.
Lines 15--17 contain \texttt{\textup{CQSBE}}, the core evolution mechanism that maintains DABNet, which is detailed in \Sec\r{the next section}.
Through this procedure, autonomous blocks evolve adaptively and remain in a \texttt{STABLE} state over the long term. When failures occur, blocks briefly enter other states and recover to the \texttt{STABLE} state via expansion, shrinking, or release, thereby preserving the overall stability of DABNet.

However, describing the evolution mechanism alone does not determine what block structure is optimal for network robustness. To guide evolution toward such configurations, DABNet's structural properties are modeled and block evolution is formalized as a robust MCN partitioning problem in the following subsection.

\subsection{Connection-Quality-Guided Partitioning}
\label{sec:pro_form}

The evolution objective of DABNet is to provide more options and alternative paths for both intra-block routing and inter-block forwarding in the presence of random ISL failures, thereby achieving a robust overlay overlay network. To this end, the composition of DABNet is first formally modeled and the evolution of autonomous blocks is formulated as a Robust MCN Partitioning (\textbf{RMP}) problem.

\subsubsection{DABNet Model}

The underlay MCN is first modeled as a graph $\mathcal{G}_{underlay} = (\mathcal{S},\mathcal{E})$, where $\mathcal{S} = \{s_1, s_2, \cdots, s_{N_s}\}$ denotes the set of satellites and $\mathcal{E} = \{e_1, e_2, \cdots, e_{N_{\mathcal{E}}} \}$ denotes the set of ISLs. Then, the whole DABNet can be represented by the combination of blocks, vagrant satellites, and faulty satellites as:
\begin{eqnarray}
    \mathcal{G}_{overlay} = (\bigcup_{k = 1}^{N_{\mathcal{B}}}   \mathcal{B}_k \bigcup \mathcal{S}_v \bigcup \mathcal{S}_f,\mathcal{E}_{IUL})
\end{eqnarray}
where $\mathcal{B}_k$ is the $k$-th block (a subgraph of $\mathcal{G}_{underlay}$), 
$\mathcal{S}_v$ are vagrant satellites, $\mathcal{S}_f$ are faulty satellites, and 
$\mathcal{E}_{IUL} \subseteq \mathcal{E}$ is the set of ISLs connecting distinct FUs.

\subsubsection{Robust MCN Partitioning Problem}

A robust DABNet must satisfy two requirements under intermittent failures: first, autonomous blocks must maintain strong internal connectivity to sustain structural stability during intra-block link fluctuations and support efficient intra-block routing; second, blocks must have a sufficient number of directionally diverse inter-unit links to provide multiple alternative paths for inter-block forwarding. Based on these requirements, the Connection Quality Score (CQS) is introduced to characterize the structural quality of autonomous blocks, defined as follows.
\begin{eqnarray}
     \psi (\mathcal{B}_k) &=& \mathop{vol}(\mathcal{B}_k) + \alpha \mathop{div}(\mathcal{B}_k) \\
    \mathop{div}(\mathcal{B}_k) &=& \dfrac{|\mathcal{E}_{IUL}(\mathcal{B}_k)|}{\left| \sum \vec{e_j} \right|}, \ e_j \in \mathcal{E}_{IUL}(\mathcal{B}_k)
\end{eqnarray}
where $\psi (\mathcal{B}_k)$ denotes the CQS of block $\mathcal{B}_k$; 
$\mathop{vol}(\mathcal{B}_k)$ denotes the block volume, defined as the sum of links of all satellites within the block, characterizing its internal connectivity; 
$\mathop{div}(\mathcal{B}_k)$ denotes the directional diversity of IULs, measuring the dispersion of the block's external connections; $\mathcal{E}_{IUL}(\mathcal{B}_k)$ is the set of IULs of block $\mathcal{B}_k$; $\vec{e_j}$ represents the direction vector of the $j$-th IUL; $|\mathcal{E}_{IUL}(\mathcal{B}_k)|$ in the numerator is the number of IULs, and $\left| \sum \vec{e_j} \right|$ in the denominator is the magnitude of the resultant vector of all IUL direction vectors---the smaller this value, the more directionally dispersed these IULs are; $\alpha$ is a weight parameter that balances internal connectivity and external connection diversity.

Accordingly, the evolution of autonomous blocks in DABNet is further formalized as the RMP problem, whose objective is to maximize the sum of CQS over all autonomous blocks subject to block size and diameter constraints. The resulting network yields the most robust DABNet. The RMP problem can be formulated as:

\noindent\textbf{~~~~Objective}
\begin{eqnarray}
     &\mathop{\textbf{max}~~~~} \sum\limits_{\mathcal{B}_k \subseteq \mathbf{B}} \psi (\mathcal{B}_k)&
\end{eqnarray}
\noindent\textbf{~~~~Subject to}
\begin{eqnarray}
     \label{eq:cst1}
    & |\mathcal{B}_k| < \dfrac{1+\epsilon}{N_{\mathcal{B}}} \sum \limits_{k = 1}^{N_{\mathcal{B}}} |\mathcal{B}_k| &\\
    \nonumber\\
    \label{eq:cst2}
    & D_{min} \leq D(\mathcal{B}_k) \leq D_{max} &
\end{eqnarray}
where $\mathbf{B}$ denotes the set of all autonomous blocks; $|\mathcal{B}_k|$ denotes the size of block $\mathcal{B}_k$, i.e., the number of satellites within the block; $\epsilon$ is a tolerance parameter controlling block size balance; $D_{\min}$ and $D_{\max}$ are the minimum and maximum allowable block diameters, respectively.
Constraint \eqref{eq:cst1} limits individual block sizes to prevent excessively large blocks from undermining partition balance. 
\udm{Constraint \eqref{eq:cst2} bounds the block diameter to control intra-block hop counts and ensure manageability. $D(\mathcal{B}_k)$ denotes the block diameter, defined as the maximum shortest-path hop count between any two nodes within the block. Larger block diameters provide richer connectivity but incur higher intra-block convergence overhead.}

\subsubsection{CQS-aware Block Evolution Algorithm}
\label{sec:alg}
The RMP problem aims to maximize the overall connection quality subject to block size and diameter constraints.
However, RMP problem is essentially a generalization of the classic max-cut problem~\cite{bulucc2016recent} and is NP-hard, making it impractical to solve exactly via global optimization in dynamic MCNs. Moreover, even if optimal solutions were available, frequently messaging in highly dynamic for global MCN partitioning would incur substantial control overhead and fail to respond to ISL fluctuations in a timely manner.
A heuristic distributed Connection-Quality-Score aware Block Evolution (CQSBE) algorithm is therefore designed, which runs locally on each autonomous block to approximately solve the RMP problem, enabling online adaptive evolution of autonomous blocks.

As shown in \alg\ref{alg:bbe}, the algorithm computes $\psi$ for each candidate merge with a neighboring vagrant satellite and selects the one yielding the highest CQS (line~6). When the block diameter exceeds $D_{\max}$ due to internal link failures, the algorithm removes the satellite with the lowest closeness centrality $\phi$ (line~8), where $\phi$ is defined as:
\begin{eqnarray}
    \phi(u) = \dfrac{|\mathcal{B}|-1}{\sum\limits_{v \in \mathcal{B},\, v \neq u} d(u,v)}
    \label{eq:phi}
\end{eqnarray}
Here, $d(u,v)$ denotes the shortest-path hop count between satellites $u$ and $v$, and $|\mathcal{B}|$ is the number of satellites in the block. \udm{This $\phi$-guided satellite removal is computed locally within each block rather than on the global graph, ensuring that the block topology remains within a low diameter range while avoiding expensive global computations.}
Since the block size that maximizes $\psi$ may differ across MCN of different scales, the algorithm introduces a parameter $D_{\max}$ to control the maximum block diameter, adapting to networks of varying sizes.
\begin{algorithm}[t!]
    \caption{Connection Quality Score aware Block Evolution Algorithm (\textbf{CQSBE})}
    \label{alg:bbe}
    \LinesNumbered 
    \KwIn{Block $\mathcal{B}_k$}
    \KwOut{Evolved block $\mathcal{B}_k^\star$ }

    \SetKwFunction{CQSBE}{CQSBE}
    \SetKwProg{Fn}{def}{\string :}{}

    \uIf{\texttt{\textup{Diameter}}($\mathcal{B}_k$)  $< D_{max}$}{
        $\mathcal{S}_{adj} \leftarrow \texttt{\textup{Adj}}(\mathcal{B}_k)$;~
        $\mathbf{B} \leftarrow \emptyset$;\tcp{block expand} 
        \For{$s_j \in \mathcal{S}_{adj} $ \&\& \texttt{\textup{State($s_j$)=VAGRANT}}} {
            $\mathcal{B}^\prime \gets \mathcal{B}_k + s_j + \mathcal{E}(s_j,\mathcal{B}^\prime)$;~
            $\mathbf{B}  \leftarrow \mathbf{B} \bigcup \mathcal{B}^\prime $;
        }
        
        $\mathcal{B}^\star_k \leftarrow \mathop{argmax}\limits_{ \mathcal{B}^\prime \in  \mathbf{B} } \psi(\mathcal{B}^\prime)$;\\

   
    }\Else{
        \tcp{block shrink} 
    $s_r \leftarrow \mathop{argmin}\limits_{s \in \mathcal{S}^\star}$   
    $\phi(s_i,\mathcal{B}_k)$;~
    $\mathcal{B}^\star_k \leftarrow \mathcal{B}_k - s_r$;
    }
    \Return{$\mathcal{B}^\star_k$}
    \end{algorithm}

\udm{
    This distributed, heuristic block evolution algorithm provides an efficient and practically realizable solution to the RMP problem. Rather than pursuing a theoretically global optimal partition, the algorithm prioritizes distributed dynamic adaptability in the face of network fluctuations.
}
Under time-varying topologies and intermittent ISLs, CQSBE enables DABNet to dynamically evolve autonomous blocks, maintaining a sufficient number of well-distributed IULs and thereby ensuring the robustness of the overlay network.

\section{Hybrid Routing over DABNet}
\label{sec:dabr}

BlockFlex designs and integrates a hybrid routing method, DABR, on top of the virtual overlay network DABNet. This routing method enables hierarchical and hybrid routing decisions within the autonomous-block-based overlay. Specifically, DABR introduces the following components: an efficient addressing strategy that mitigates the impact of Earth's rotation and mobile node position uncertainty on the addressing process; an n-step backward acknowledgment signaling-based protection routing mechanism to ensure routing resilience against dead-end issues in inter-block routing; an optimal source satellite selection algorithm to improve end-to-end performance stability under the latency and jitter caused by enhanced inter-block connectivity; and finally, a dual-mode forwarding switching mechanism that reconciles the incompatible forwarding modes between stateless geographic routing and the stateful protection routing and source satellite selection mechanisms.

\subsection{Core Routing Paradigm}
\label{sec:inter}

\begin{figure}[t!]
    \centering
    \includegraphics[width=0.95\linewidth]{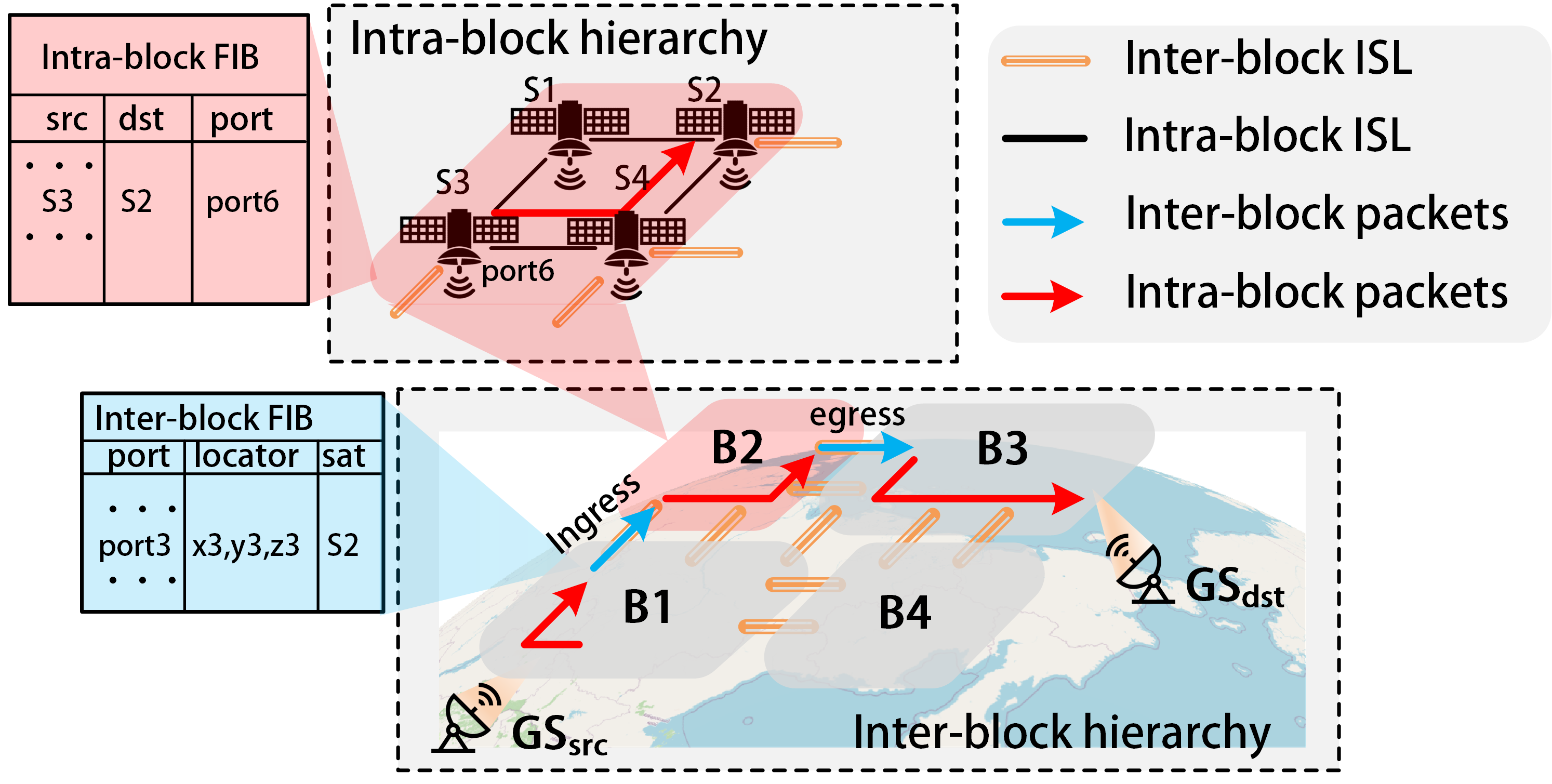}
    \caption{The core routing paradigm of DABR.}
    \label{fig:dabr}
    \vspace{-1em}
\end{figure}

\fig\ref{fig:dabr} illustrates the core routing paradigm of DABR, which consists of an intra-block routing hierarchy and an inter-block routing hierarchy. Specifically, intra-block routing employs a LSA-based mechanism to maintain a shortest-path-based Forwarding Information Base (FIB) at each satellite (shown in red in \fig\ref{fig:dabr}). This FIB specifies the next-hop port for routing between any two satellites within the same autonomous block, thereby providing reachability-guaranteed internal routing and confining convergence scope within the current block. Inter-block routing relies on another FIB maintained at all satellite nodes that records the IUL state information of the current block (shown in blue in \fig\ref{fig:dabr}), whose information is also synchronized via the intra-block routing's LSA mechanism. When a packet is forwarded to any satellite within the current block, the satellite performs forwarding decisions using both FIBs.

\subsubsection{Inter-block Routing} 

\begin{figure}[t!]
    \centering
    \subfloat[vagrant satellite forwarding]{
        \includegraphics[width=0.5\linewidth]{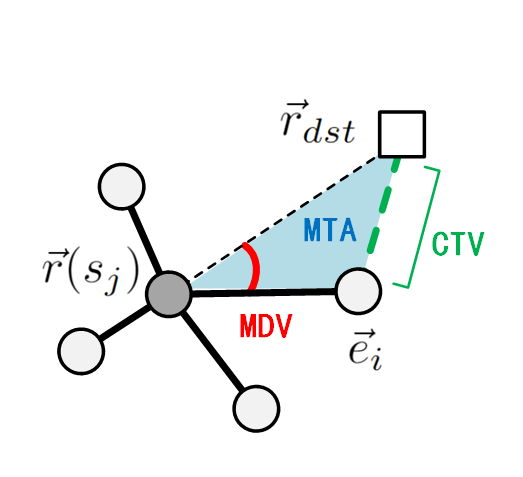}
    }\subfloat[\udm{anonymous block forwarding}]{
        \includegraphics[width=0.5\linewidth]{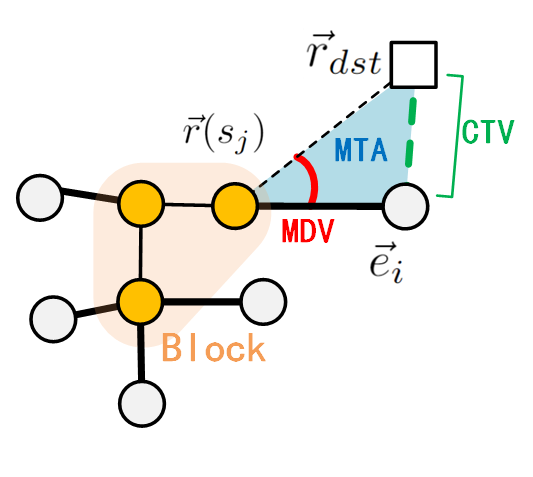}
    }
        \caption{Typical geographic packet forwarding methods. The circle represents satellites while square stands for GSes.
        }
        \label{fig:geor}
\end{figure}

Similarly to geographic methods for satellite network routing\cite{navas1997geocast,henderson2000distributed,tsunoda2006geographical,xu2022spatial,lai2021orbitcast,liu2022geographic}, DABR uses real-time Global Navigation Satellite System service \cite{kaplan2017understanding} to perform geographic forwarding between blocks and vagrant satellites. These methods share a common principle: selecting the egress link and completing the forwarding decision based on the geometric relationship formed by the positions of the current satellite node, the candidate next-hop satellite node, and the destination node. 
DABR unifies closer to target vertex forwarding, minimum deviation vertex forwarding, and minimum triangle area forwarding into inter-block routing.
The forwarding principle is illustrated in \fig\ref{fig:geor}, where (a) shows the triangular geometric criterion used by vagrant satellites for packet forwarding, and (b) shows that used by autonomous blocks. Specifically, when the current autonomous block or vagrant satellite selects the next-hop egress link $e_{egress}$, the three forwarding computation methods are as follows:

\textbf{Closer to Target Vertex ({\textcolor[rgb]{0,0.7,0}{CTV}})}:
\begin{eqnarray}
    \label{eq:ctv}
	e_{egress} =\mathop{argmin}\limits_{e_i \in \mathcal{E}}  \{ | \vec{e}_i + \vec{r}(s_j) - \vec{r}_{dst}| \}
\end{eqnarray}

\textbf{Minimum Deviation Vertex ({\textcolor[rgb]{1,0,0}{MDV}})}:
\begin{eqnarray}
    \label{eq:mdv}
	e_{egress}=\mathop{argmin}\limits_{e_i \in \mathcal{E}}  \{ \dfrac{\vec{e_i} \cdot (\vec{r}_{dst}- \vec{r}(s_j))}{|\vec{e}_i||\vec{r}_{dst}-\vec{r}(s_j) |}\}
\end{eqnarray}

\textbf{Minimum Triangle Area ({\textcolor[rgb]{0,0,0.8}{MTA}})}:
\begin{eqnarray}
    \label{eq:ctmd}
	e_{egress} = \mathop{argmax}\limits_{e_i \in \mathcal{E}}  \{ \dfrac{\vec{e_i} \cdot (\vec{r}_{dst}- \vec{r}(s_j))}{|\vec{e}_i||\vec{r}_{dst}-\vec{r}(s_j) | | \vec{e}_i + \vec{r}(s_j) - \vec{r}_{dst}| }\}
\end{eqnarray}
Here, $\mathcal{E}$ denotes the set of all possible egress ISLs in the current FU; $s_j$ and $\vec{r}(s_j)$ represent the current autonomous block or vagrant satellite and its locator, respectively; $g_{dst}$ and $\vec{r}_{dst}$ denote the destination node and its locator, respectively. DABR uses MTA as the default forwarding method, as it achieves a favorable balance between CTV and MDV.

\subsubsection{Intra-block Routing}
\label{sec:intra}

Geographic routing is convergence-free and offers high forwarding efficiency; however, it may still encounter dead-end issues near the destination due to local minima \cite{henderson2000distributed,wood2001internetworking}, and this problem becomes more severe as the number of failed links increases. To address this, DABR introduces a LSA-based routing method within autonomous blocks, which cooperates with inter-block geographic routing to complete the full packet forwarding process. Specifically, each autonomous block periodically updates its intra-block FIB via a LSA mechanism, which records the shortest paths between any two satellites within the block. 
When a packet enters a block through a satellite $s_{ingress}$, the satellite $s_{ingress}$ first checks whether the destination GS is attached to any satellite $s_{egress}$ within the block. If so, the packet is routed to $s_{egress}$ and forwarded to the destination GS via a GSL; otherwise, it is routed to the appropriate egress satellite to enter the next autonomous block. When an IUL failure occurs, the autonomous block first recovers to a stable and connected topology through its adaptive evolution mechanism, then triggers a LSA to update the FIB and complete routing re-convergence. 
Since the block size is constrained, the overhead of LSA, FIB updates, and convergence is all confined within the block boundary, thereby forming an isolated convergence domain \cite{rfc4271} that prevents local fluctuations from flooding the entire network.

\subsection{Decoupled and Rotation-Invariant Addressing}
\label{sec:addressing}

Beyond ISL failures, the dynamics of MCNs also manifest in the intermittent connectivity between LEO satellites and edge nodes, including GSes, user receivers, civil aviation aircraft, and even remote sensing satellites. The frequent link handover between these nodes and satellites, compounded by the uncertainties introduced by Earth's rotation, pose significant challenges in the addressing process. 
Although existing studies~\cite{pan2019opspf,ali1999predicting,li2025small,liu2025leo} have attempted to predict the destination satellite currently adjacent to a destination node based on orbit forecasting, such methods are ill-suited for high-speed mobile nodes, because the source node typically cannot obtain the real-time position of the destination node and thus cannot accurately predict its currently adjacent satellite, ultimately leading to addressing failures. To this end, DABR proposes two key addressing design principles to cope with the aforementioned dynamics.

\subsubsection{Locator-Identifier Decoupling}
\begin{figure}[t!]
    \centering
    \includegraphics[width=0.85\linewidth]{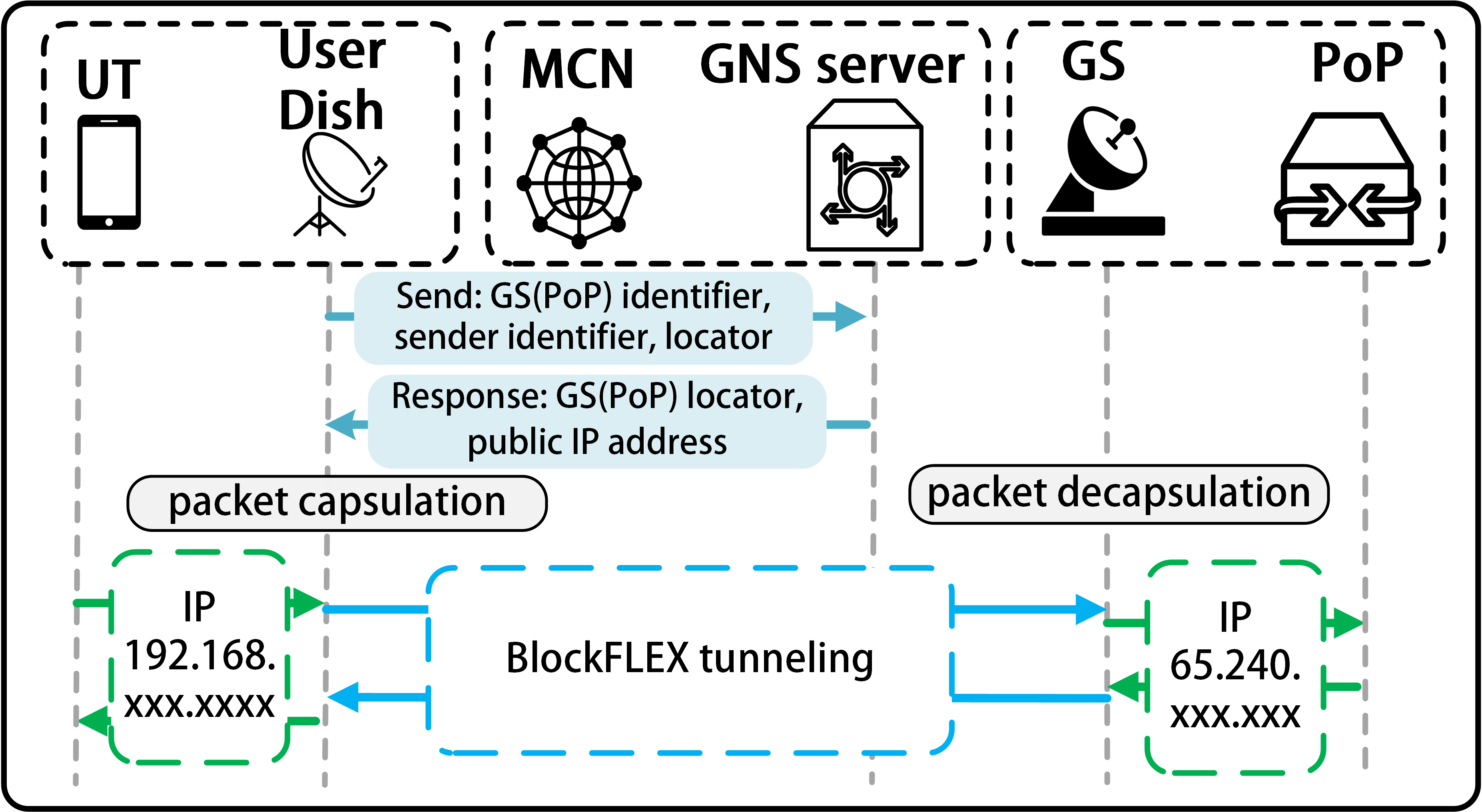}
    \caption{GNS signaling procedures.}
    \label{fig:gns}
\end{figure}

End-to-end addressing in highly dynamic network requires a clear separation between what a node is and where it is currently located \cite{clark2018designing,venkataramani2014mobilityfirst,zhang2017supporting}. 
DABR decouples persistent, location-independent identifiers from ephemeral, topology-dependent locators, as is shown in \fig\ref{fig:gns}. 
Each edge node such as a user dish or GS is assigned a stable identifier. 
Before sending a packet, the source node queries a distributed Global Name Service (GNS) \cite{venkataramani2014mobilityfirst} 
to resolve the destination’s identifier into its current addressing locator.
Edge nodes only update their locators at the GNS when they change their topological attachment satellite. 
Meanwhile, routing simply forwards packets using the current locator carried in the packet header, remaining unaware of end node movements. 

\subsubsection{Rotation‑Invariant Addressing via ECI Coordinate}

The high relative dynamics between satellites and ground, coupled with Earth’s rotation, pose significant challenges for traditional ground‑referenced addressing schemes.
Unlike approaches that rely on complex mapping mechanisms to address such coupling \cite{lai2021orbitcast,li2024stable}, this complex mapping can be resolved by adopting a 3‑dimensional Earth‑Centered Inertial (ECI) coordinate system \cite{rizzi2004relativity} as the addressing space, in which both satellite and ground‑station positions are expressed within the same non‑rotating reference frame, thereby decoupling addressing from Earth’s rotation.
Every node uses the high-precision timing and positional information provided by GNSS to convert its own location and the destination's location into ECI coordinates.

The addressing scheme in DABR offloads mobility management to a dedicated naming service, which reduces control-plane overhead and enables seamless handovers. Additionally, its support for interoperability with existing terrestrial protocols allows BlockFlex to function as an efficient tunneling overlay.
Its decoupled design inherently accommodates heterogeneous edge devices with diverse mobility (e.g., maritime vessels, high-altitude platforms, or remote-sensing satellites) through updates only to the GNS, without modifying core routing logic. 
Furthermore, the continuous 3-dimensional ECI coordinate system provides a rotation-invariant addressing space that natively supports multi-shell mega-constellation networks \cite{starlink2,StarlinkTechOverview}, eliminating domain-specific coordinate conversions. This approach not only resolves dynamic edge-network interactions and Earth-rotation distortions, but also ensures the long‑term scalability for future network expansion.

\begin{figure}[t!]
	\centering
        \includegraphics[width=1\linewidth]{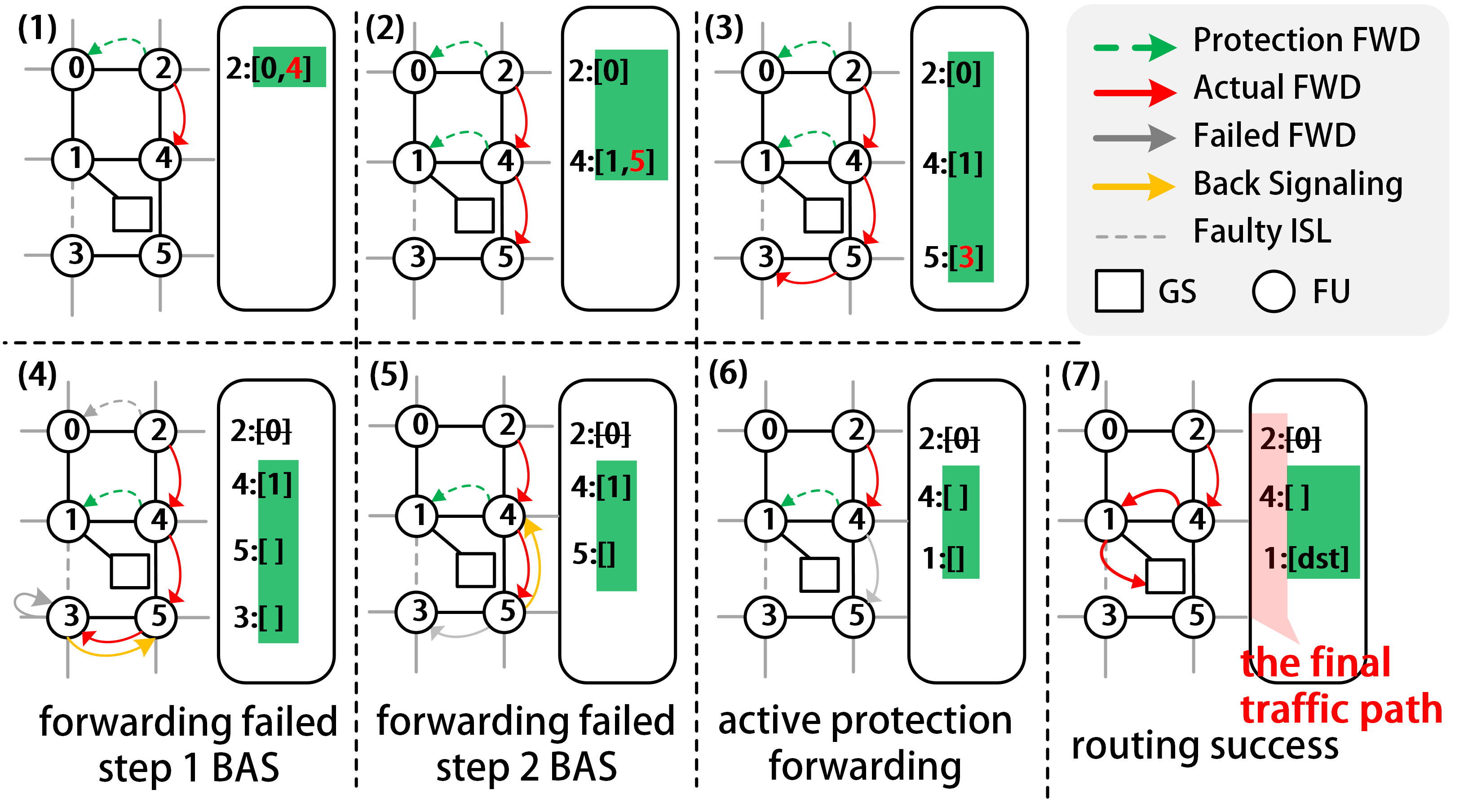}
        \vspace{-1em}
		\caption{The n‑step backward acknowledgment signaling mechanism for resilient routing. For clarity, forwarding units, whether vagrant satellites or blocks, are represented by circles.
        The white panel in the right of sub-figures present a global view of the routing.}
        \vspace{-1em}
		\label{fig:nbas}
\end{figure}

\subsection{Routing Failure Recovery}
\label{sec:nbas}

While DABNet confines most ISL failures within blocks, inter‑block geographic forwarding can still encounter local minima or dead-end disruptions\cite{henderson2000distributed,eISL}. 
To address this, a resilient routing method is proposed based on the n-step Backward Acknowledgment Signaling (nBAS) mechanism. 
Each FU along the path of dataflow $I_D$ maintains a protection forwarding stack (PFS) $\mathcal{P}[I_D]$, which stores all feasible next-hop options. If routing fails, a backward signaling process is triggered recursively to retrieve cached alternatives and activate backup forwardings.
\fig\ref{fig:nbas} illustrates the case of nBAS process, which operates in three sequential steps:

\noindent\ding{182}~\textbf{PFS caching:} As the dataflow travels (e.g., $U_2 \rightarrow U_4 \rightarrow U_5$), each FU computes and ranks all possible egress links by a geographic matching score (\eqt\ref{eq:ctv}–\ref{eq:ctmd}), storing them in the PFS, not just the optimal next hop.

\noindent\ding{183}~\textbf{Failure handling:} When a FU $U_3$ fails to forward packet in $I_D$, it checks its PFS. If empty, a \texttt{NACK} is sent upstream to $U_{5}$. If that PFS is also empty, the signal continues to $U_{4}$ (two-hop fallback).

\noindent\ding{184}~\textbf{Protection forwarding active:} Once a FU with a non-empty PFS (e.g., $U_{4}$) is reached, it pops a backup forwarding (e.g., $U_4 \rightarrow U_1$), confirms adjacency to the destination GS $g_{dst}$, and redirects traffic via $U_1 \rightarrow g_{dst}$.

Unlike approaches that propagate link-state updates network-wide upon ISL failures\cite{lai2023achieving,du2025multi,liu2025leo}, nBAS operates in a fully distributed manner: each FU makes independent decisions based on its local PFS and recursively propagates \texttt{NACK} signals upstream. By leveraging limited-step backtracking without requiring global topology information, nBAS implements a form of distributed depth-first search that achieves efficient path recovery with minimal overhead.

\begin{figure}[t!]        
    \centering
    \includegraphics[width=0.7\linewidth]{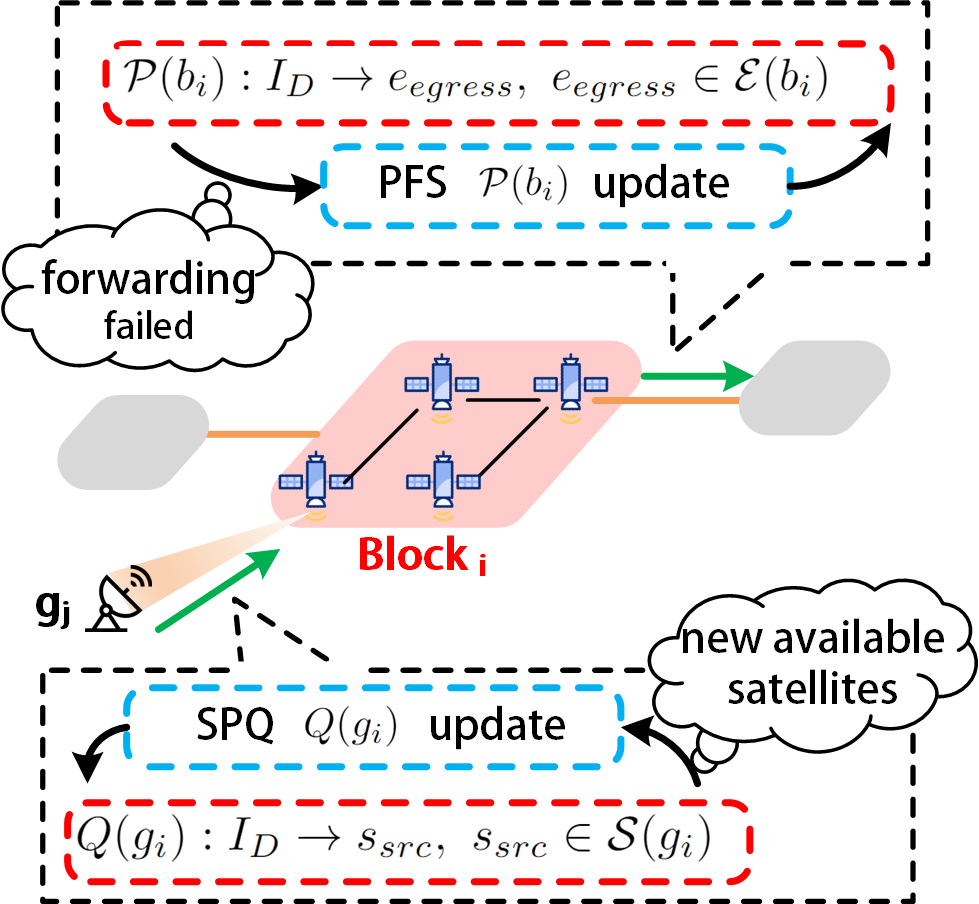}
    \caption{Forwarding mode switching.}
    \label{fig:dual}
  \end{figure}

\begin{figure*}[t!]
    \centering
    \includegraphics[width=0.95\linewidth]{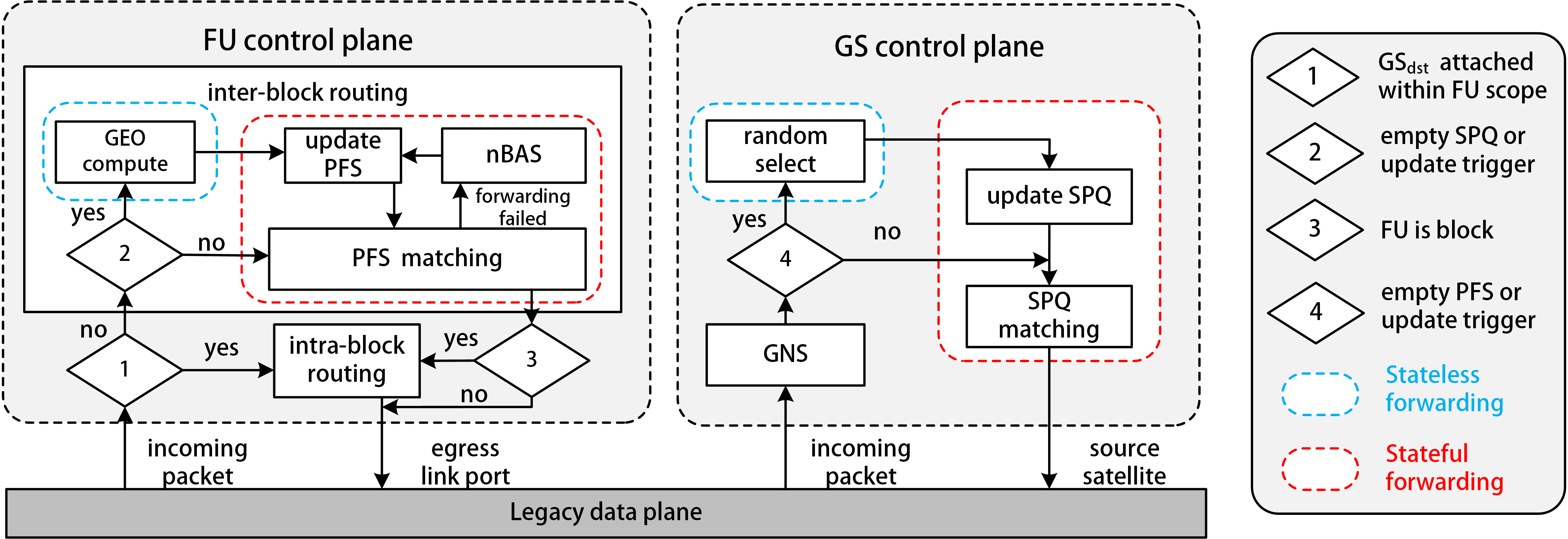}
                \caption{BlockFlex control plane overlay.}
                \label{fig:overlay}
  \end{figure*}
  
\subsection{Routing Path Optimization}
\label{sec:os3}

The enhanced connectivity between blocks increases path diversity, which may also lead to higher latency and jitter. While selecting an appropriate source or destination satellite \cite{liu2024efficient,zhang2022enabling} could significantly mitigate these effects, such an approach typically requires centralized control‑plane intelligence. 
Substantial performance gains can still be achieved through a fully distributed strategy that focuses solely on source satellite selection, leveraging feedback from ongoing traffic. 
The Optimal Source Satellite Selection (OS3) algorithm is therefore introduced—a lightweight, fully distributed mechanism that dynamically selects the optimal ingress satellite based on measured latency, thereby stabilizing end‑to‑end performance without centralized coordination.

In OS3, each GS maintains a Source-satellite Priority Queue (SPQ) $\mathcal{Q}[I_D]$ for each dataflow $I_D$ to store the optimal source satellites, which are prioritized according to the latest round-trip time (RTT) of the $I_D$. The RTTs are collected by randomly selecting source satellites during the initial phase of the dataflow, and $\mathcal{Q}[I_D]$ will be dynamically updated as available satellites change.
 As the dataflow continues, it will gradually converge toward the source satellite with the optimal RTT.

\subsection{System Integration via Forwarding Mode Switching}
\label{sec:dual}
Inter-block geographic routing (\Sec\ref{sec:inter}) operates in a stateless manner, making forwarding decisions per-packet based on geographic computation. In contrast, mechanisms such as nBAS resilient routing (\Sec\ref{sec:nbas}) and the OS3 source satellite selection policy (\Sec\ref{sec:os3}) rely on per-flow state, i.e., 
the cached context PFS and SPQ of dataflow $I_D$, which follow a stateful forwarding model.
To handle this incompatibility issue, 
DABR introduces a Forwarding Mode Switching (FMS) mechanism that dynamically alternates between stateless and stateful operation in forwarding units and GSes.

As illustrated in \fig\ref{fig:dual},
During stateless mode ({\color{blue}{blue dash line}}), the GS detects a newly available satellite $s_\text{src}$, triggering the update of the SPQ $\mathcal{Q}$. Meanwhile, the blocks refresh the protection forwarding stack $\mathcal{P}$ periodically based on the latest geographic forwarding decisions.

Once $\mathcal{P}$ and $\mathcal{Q}$ are updated, all nodes switch to stateful mode ({\color{red}{red dash line}}). Forwarding units $b_i$ perform lookups in $\mathcal{P}(b_i): I_D \rightarrow e_{egress}$ to map a flow identifier $I_D$ to the corresponding egress IUL $e_{egress}$ or, if the destination GS is attached to a satellite within the block, to a GSL. Ground stations $g_j$ use $\mathcal{Q}(g_j): I_D \rightarrow s_{src}$ to select the appropriate source satellite $s_{src}$ for each flow $I_D$.
This dual-mode switching approach not only resolves the incompatibility between stateless geographic routing and stateful mechanisms such as nBAS and OS3, but more importantly, it transforms forwarding decisions from per-packet geometric computation to event-triggered updates. Packets are subsequently forwarded via cached table lookups rather than repeated on‑the‑fly calculations, substantially reducing computational overhead.

\fig\ref{fig:overlay} illustrates the complete control plane of DABR, which integrates the mechanisms described in the preceding sections and operates in a fully distributed manner.
At each edge node (e.g., a GS or user dish), the control logic invokes the GNS and selects the optimal source satellite for uplink transmission.
Within each forwarding unit, the control logic determines the appropriate egress link for every incoming packet, whether it be a GSL for downlink transmission or an ISL for satellite forwarding. 
Both the forwarding unit and GS control planes are built upon a unified data plane.

\section{Evaluation}
\label{sec:exp}
The BlockFlex prototype is implemented in the open‑source simulator SNK \cite{snk}, which supports large‑scale constellation simulation and network performance evaluation.
All the simulations are executed on a hardware platform featuring an Intel Core i7-10700K processor, 32GB RAM.
The performance of BlockFlex is evaluated and compared against several existing MCN routing schemes across three key aspects: robustness (\Sec\ref{exp:sur}), resiliency (\Sec\ref{exp:res}), and efficiency (\Sec\ref{exp:eff}).
The experimental setup is described as follows:

\noindent\ding{117}~\textbf{Constellations:} Experiments use two operational constellations: Starlink \cite{starlink} (1584 satellites in 72 planes at 550 km) and OneWeb \cite{oneweb} (588 satellites in 12 planes). All constellations adopt the +Grid ISL topology \cite{bhattacherjee2019network}.

\noindent\ding{117}~\textbf{Network failure model:}
To stress‑test the architecture under extreme but bounded disruption, random and unpredictable ISL failures are injected uniformly across the network. Specifically, each ISL has a fixed probability of being marked unavailable in each time step, and these probabilities are uniformly distributed across the entire network. By adjusting this probability, varying degrees of network impairment can be simulated and the performance of each scheme under different failure rates can be evaluated. 

\noindent\ding{117}~\textbf{Traffic scheme:}
Over 100 major cities and islands are selected as ground station locations. Traffic demands are generated based on a weighted average of Gross Domestic Product and population \cite{ciesingpwv4}, where the weight determines the probability of traffic generation between pairs. 
During a simulation period of $2\times 10^3$ seconds, more than 1000 GS pairs are generated as sources and destinations to simulate massive traffic flows.

\noindent\ding{117}~\textbf{Routing and resilience schemes:}
SHORT \cite{li2024stable} and OrbitCast \cite{lai2021orbitcast} are implemented based on their respective original papers as baselines. 
These schemes forward packets in a fundamentally stateless manner based on geographic information: SHORT utilizes a \textcolor[rgb]{1,0,0}{MDV}, while OrbitCast employs a \textcolor[rgb]{0,0.7,0}{CTV}. The Location Guided Protection (LGR) mechanism \cite{lai2023achieving} is also incorporated into the above routing schemes, which selects the next best available egress link upon a forwarding failure to preserve packet forwarding.
For the OSPF\cite{rfc5340} and AODV\cite{perkins2003rfc3561} protocols, routes are computed using Dijkstra's algorithm via the \texttt{networkx} tool \cite{hagberg2008exploring}, and corresponding costs are derived by incorporating their routing characteristics with the dynamic network topology.

\subsection{Robustness of DABNet Overlay}
\label{exp:sur}

As the overlay network atop the MCN underlay, DABNet's robustness refers to its ability to maintain a stable routing topology under network fluctuations. 
Accordingly, three key metrics are quantified under 30\% ISL failures with block diameters $D_{min}=2$ and $D_{max}=4$: the frequency of IUL changes, the number of vagrant satellites, and the average node degree of autonomous blocks and vagrant satellites. 
These metrics collectively reflect the robustness of DABNet. The overlay schemes compared include:
1) \textbf{BASIC:} Conventional satellite networks,
2) \textbf{STATIC:} DABNet without evolutionary adaptation after initialization,
3) \textbf{RANDOM:} DABNet employing random evolution strategies, and
4) \textbf{CQSBE:} DABNet evolved using the CQSBE algorithm (\Sec\ref{sec:alg}).

\begin{figure}[t!]
    \begin{center}
    \subfloat[OneWeb's IUL changes.]
    {   
        \includegraphics[width=0.5\linewidth]{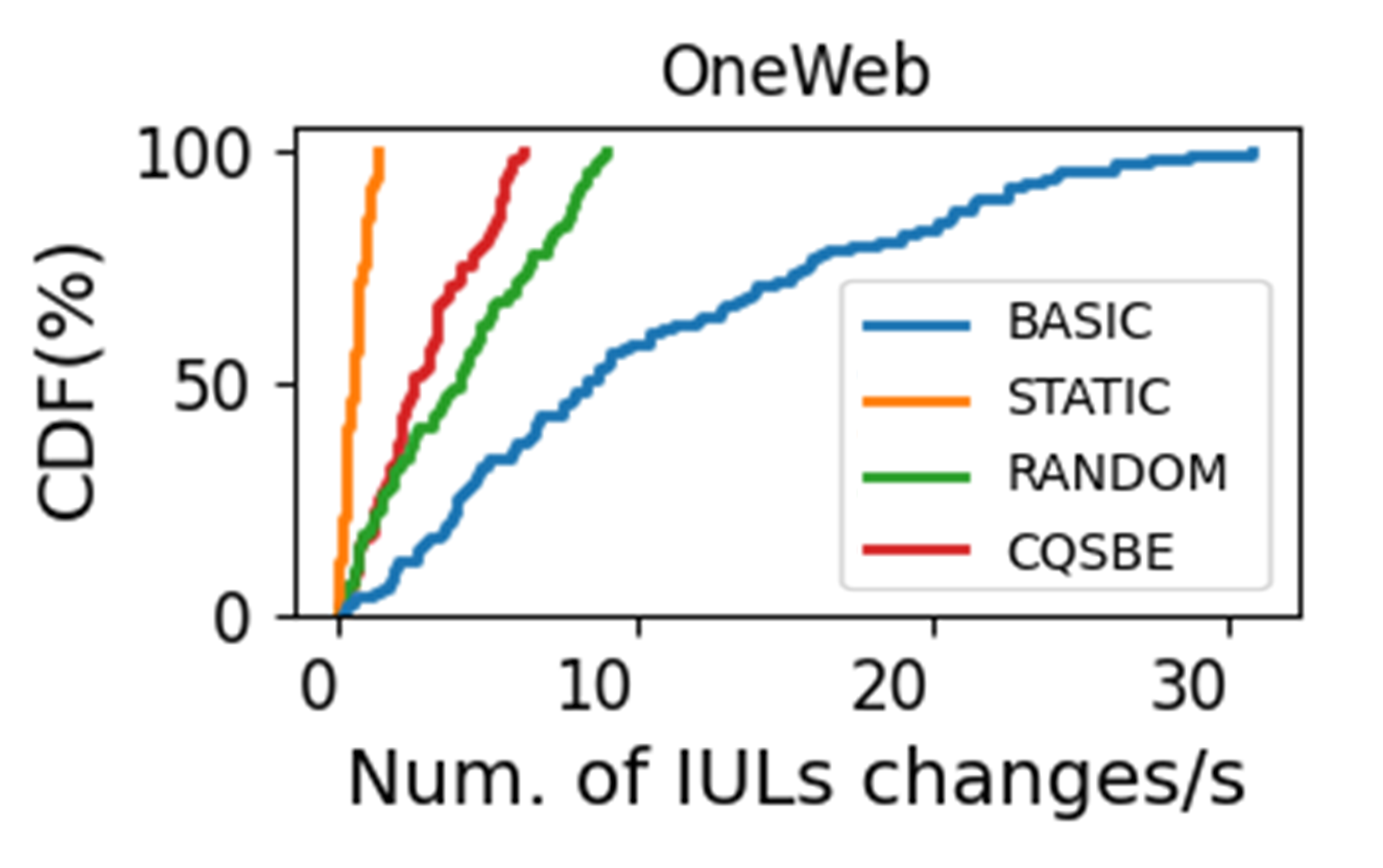}
    }\subfloat[Starlink's IUL changes.]
    {   
        \includegraphics[width=0.5\linewidth]{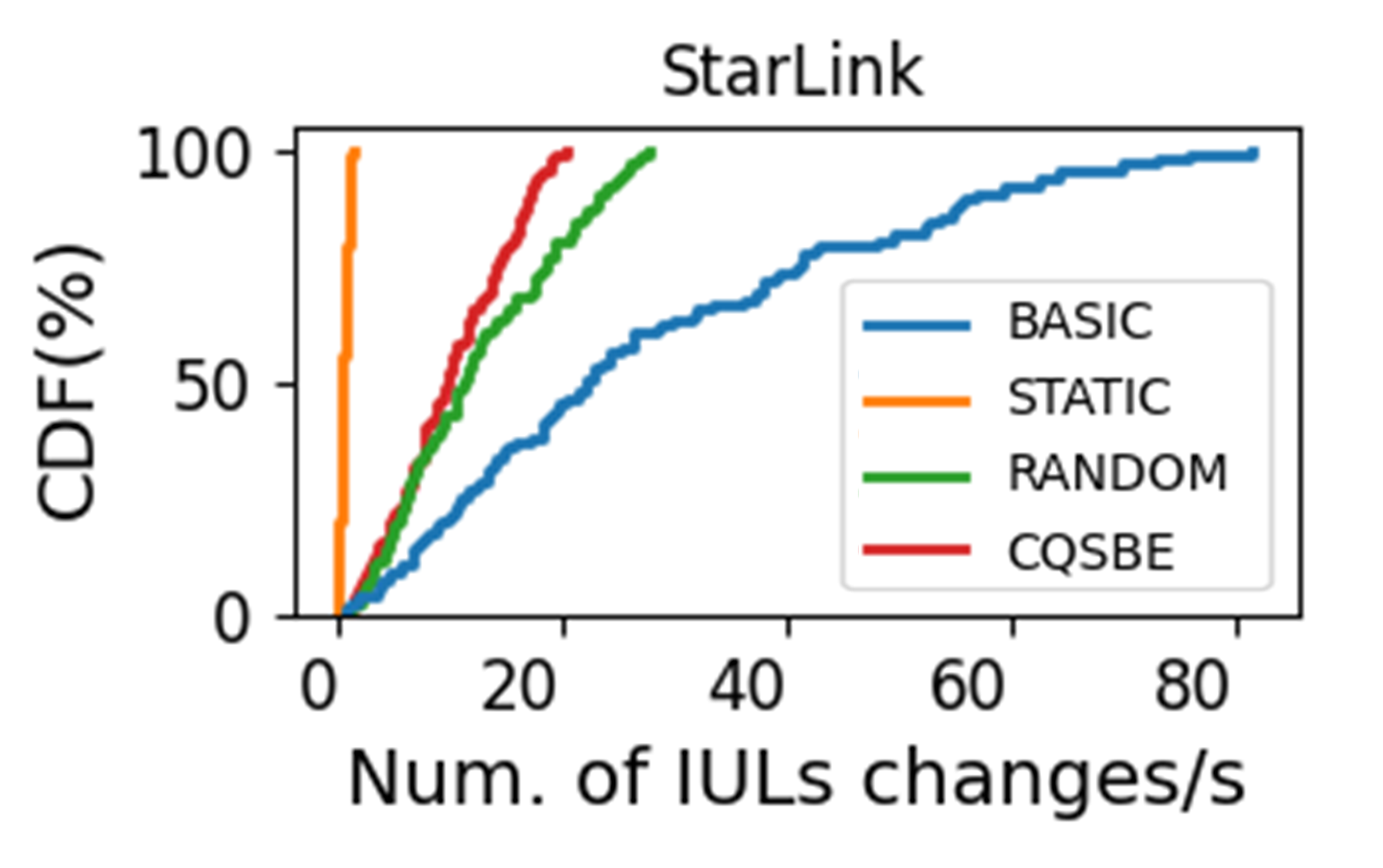}
    }
    \vspace{-1em}
    
    \subfloat[Avg. degree of FUs.]
    {
    \includegraphics[width=0.5\linewidth]{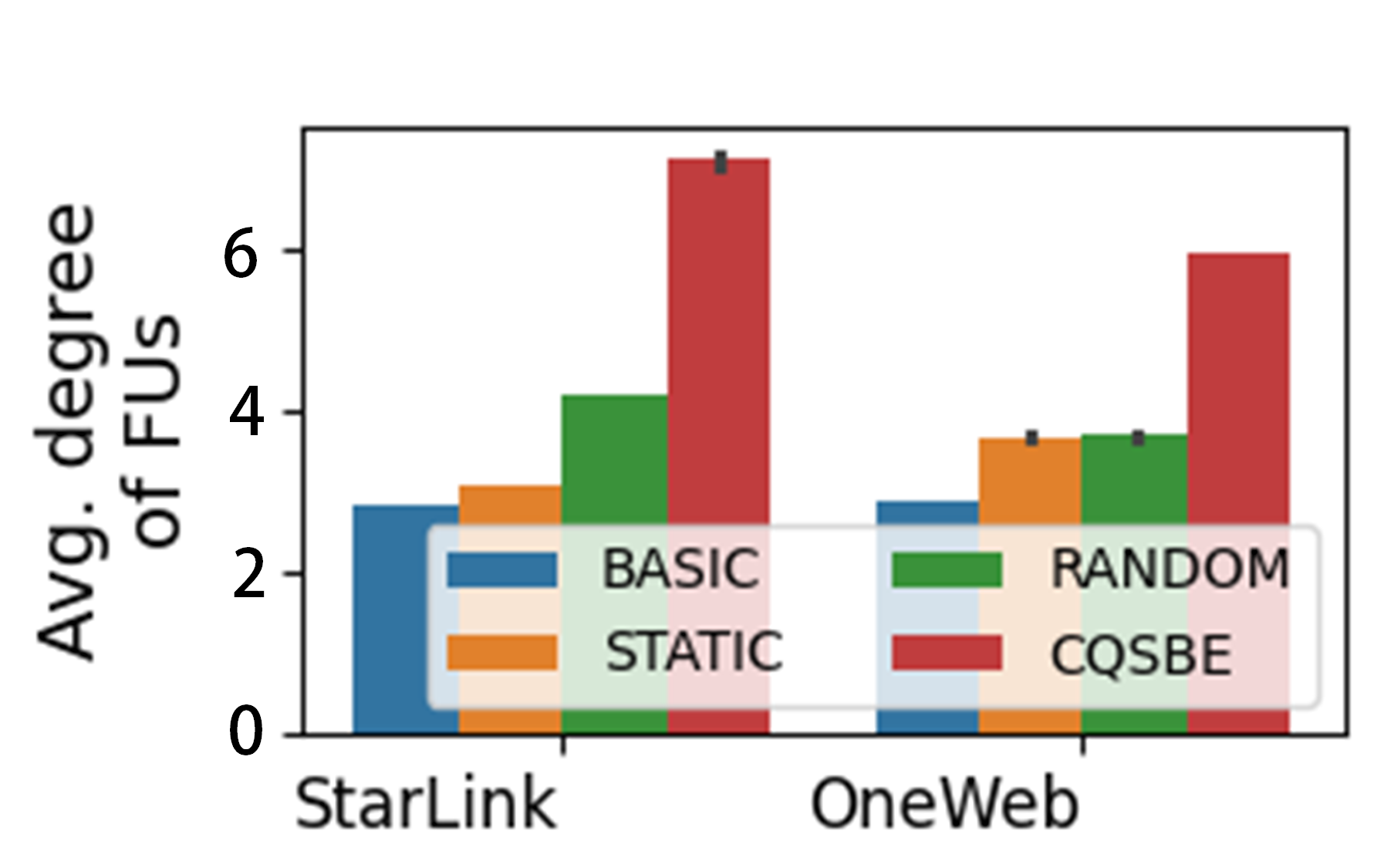}
    }\subfloat[Num. of vagrant satellites.]
    {
    \includegraphics[width=0.5\linewidth]{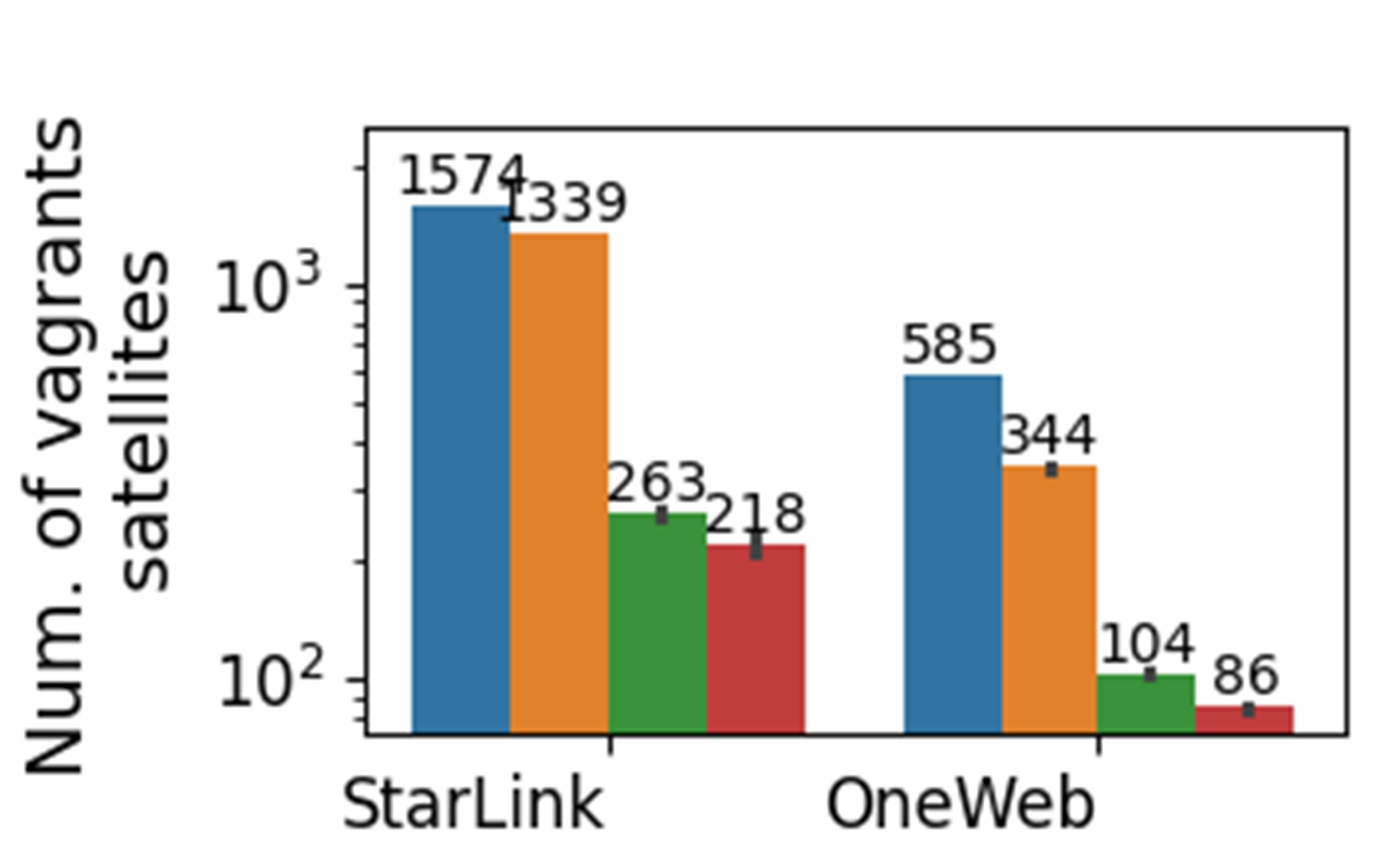}
    }

\end{center}
    \caption{Network robustness of different overlay schemes.} 
    \vspace{-1em}
       \label{fig:sur}
   \end{figure}

\fig\ref{fig:sur} (a) and (b) illustrate the distribution of IUL changes under different network maintenance schemes. 
A lower IUL variation value indicates greater topology stability, which corresponds to better network robustness. 
As shown, the BASIC scheme exhibits the highest degree of topological variation, since any link failure triggers changes in the topology between forwarding units (i.e., satellites), placing a substantial burden on maintaining normal network operation. 
The static DABNet approach achieves the lowest dynamics, as it only updates the topology view when all links within a block have failed; however, this comes at the cost of reduced connectivity over time.
By imposing more coordinated constraints on block expansion and shrinkage, the CQSBE algorithm maintains lower variation in connectivity between FUs compared to random evolution, thereby enabling DABNet to evolve with higher stability.

\fig\ref{fig:sur}(c) shows the average forwarding unit degree, where higher values indicate better connectivity. \fig\ref{fig:sur}(d) presents the number of vagrant satellites, where a lower count is preferable due to their limited IULs compared to blocks. Together, these metrics reflect the overall connectivity among forwarding units in the network.
In the BASIC scheme, all FUs are vagrant satellites. Under the +Grid connection pattern, each satellite connects to at most four ISLs, resulting in the lowest average degree. 
The static scheme, lacking a block evolution mechanism, gradually loses FU connectivity as failures accumulate, eventually approaching the performance of the BASIC scheme. The random scheme fails to converge due to the SMP problem.
In contrast, the CQSBE scheme preserves the highest connectivity under a volatile MCN, reflected in a higher average FU degree and fewer vagrant satellites. This is because the CQSBE algorithm preferentially selects vagrant satellites with higher connection quality scores, encouraging blocks to interlock in a more complementary manner. As a result, the network becomes densely populated with well-connected blocks, thereby strengthening overall connectivity among forwarding units.


%

\begin{figure}[t!]
    \centering
    \subfloat[BASIC]{
        \includegraphics[width=0.4\linewidth]{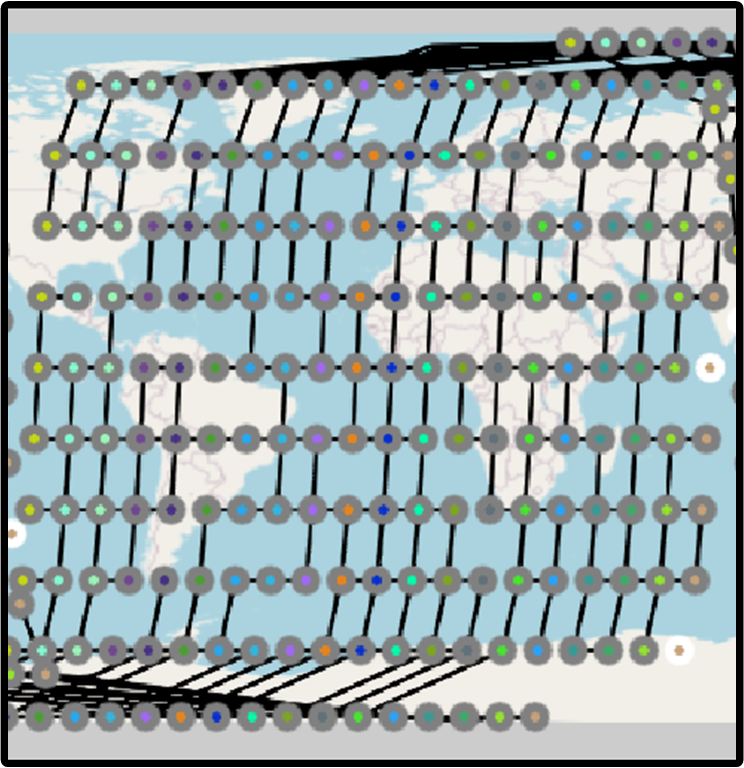}
    }
    \subfloat[STATIC]{
        \includegraphics[width=0.41\linewidth]{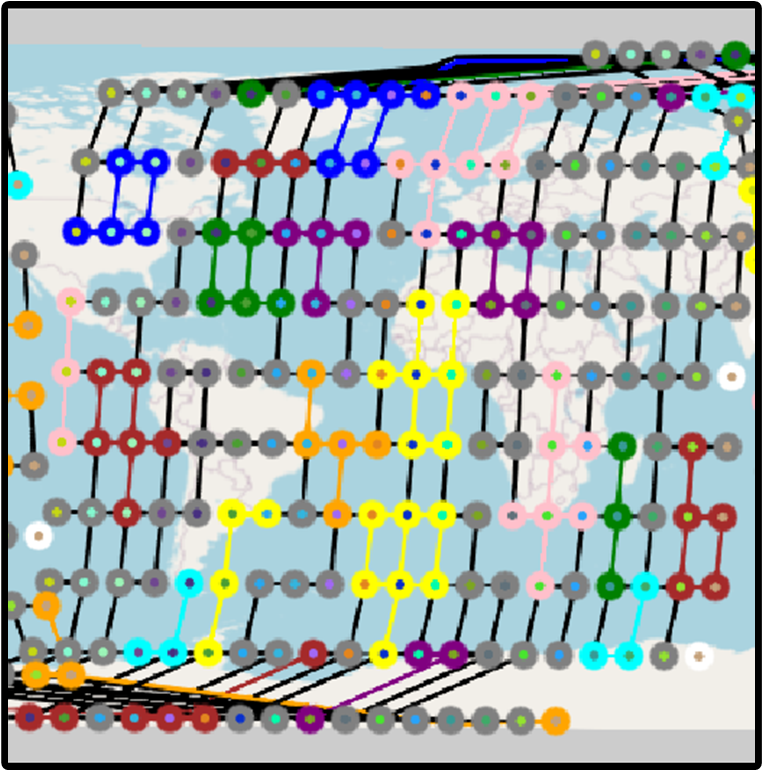}
    }
    
    \subfloat[RANDOM]{
        \includegraphics[width=0.4\linewidth]{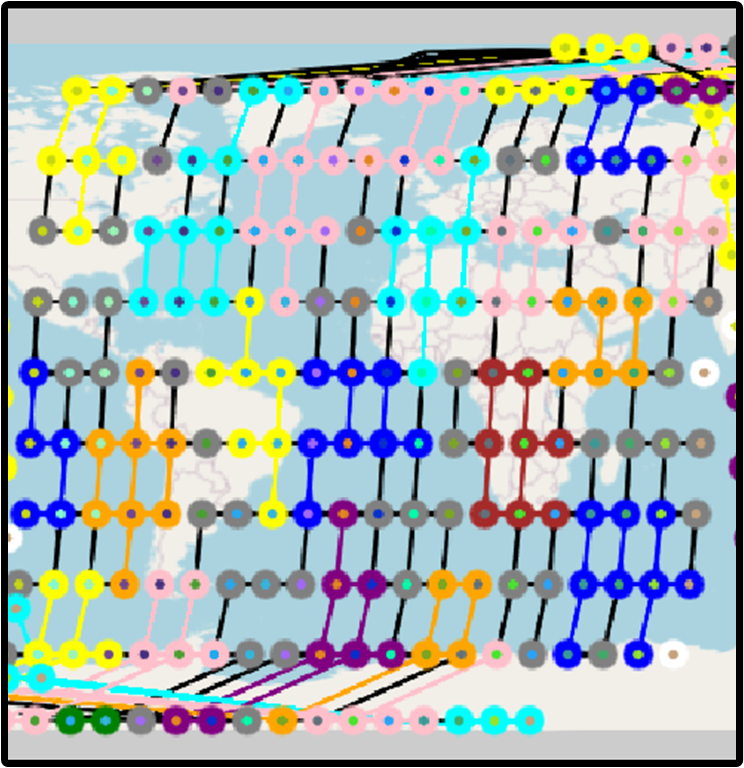}
    }
    \subfloat[CQSBE]{
        \includegraphics[width=0.41\linewidth]{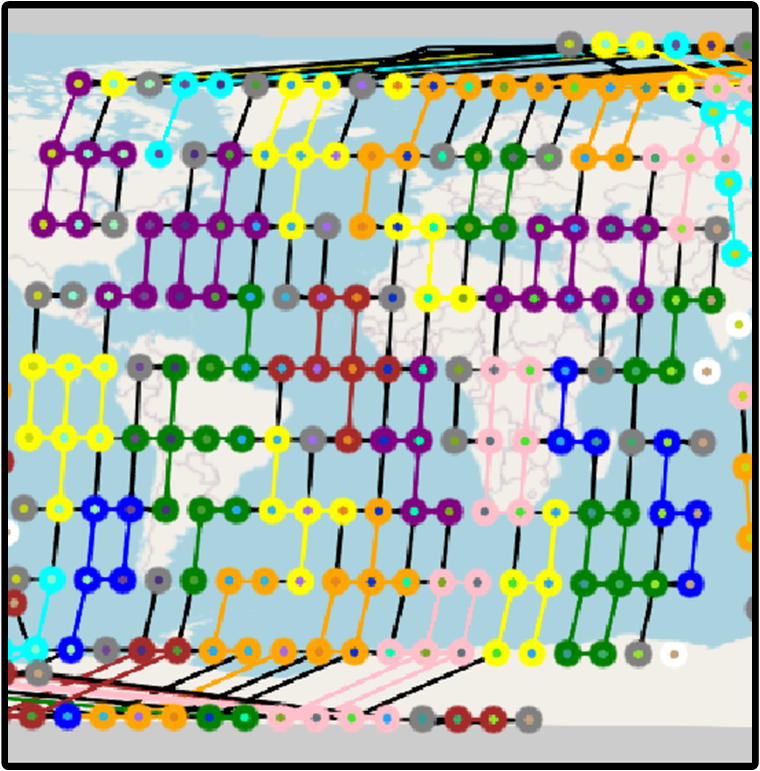}
    }

\caption{Evolution of autonomous blocks under different overlay network strategies in OneWeb} 
   \label{fig:cqsbe}
\end{figure}

\fig\ref{fig:cqsbe} illustrates the distribution of autonomous blocks under different overlay schemes in the OneWeb constellation. Evidently, under a large number of faulty satellites (white nodes), DABNet with the CQSBE method exhibits tighter block connectivity with fewer vagrant satellites, demonstrating more robust evolution and superior satellite partitioning, which further mitigates the impact of underlay network fluctuations on routing.

\begin{table*}[htbp]
	\caption{Reachability of different routing schemes. }
	\label{tab:reach}
	\centering
	\renewcommand{\arraystretch}{1.3}  
	\setlength{\tabcolsep}{2pt}
	\scalebox{0.95}{
		\begin{tabular}{c|llll|llll}
			\toprule[2pt]

			\multicolumn{1}{c|}{\multirow{2}{*}{\diagbox{Routing schemes}{Failure ratio}}} & \multicolumn{4}{c|}{Starlink} & \multicolumn{4}{c}{OneWeb}   \\ \cline{2-9}
			\multicolumn{1}{c|}{}          & \multicolumn{1}{c}{0\%}       & \multicolumn{1}{c}{10\%}      & \multicolumn{1}{c}{20\%}      & \multicolumn{1}{c|}{30\%}    & \multicolumn{1}{c}{0\%}       & \multicolumn{1}{c}{10\%}      & \multicolumn{1}{c}{20\%}      & \multicolumn{1}{c}{30\%}       \\ \midrule
			OrbitCast+LGR\cite{lai2021orbitcast}         & 94.54 ($\pm$1.10)             & 77.27 ($\pm$2.01)             & 62.62($\pm$2.63)              & 46.13 ($\pm$3.07)             & 97.21 ($\pm$0.48)             & 81.45 ($\pm$3.15)             & 64.94 ($\pm$2.77)             & 45.70($\pm 1.54$)              \\
			SHORT+LGR\cite{li2024stable}                 & 97.55 ($\pm$0.89)             & 79.09 ($\pm$2.62)             & 65.55 ($\pm$5.19)             & 49.01 ($\pm$2.68)             & 97.70 ($\pm$0.72)             & 82.29 ($\pm$2.44)             & 64.62 ($\pm$3.49)             & 44.98 ($\pm$1.85)              \\
			\midrule
			DABR(MDV)                                & 93.85 ($\pm$1.01)             & 84.15 ($\pm$1.77)             & 73.64($\pm$2.44)              & 60.10 ($\pm$2.24)             & 90.83 ($\pm$1.43)             & 79.77 ($\pm$1.99)             & 71.08 ($\pm$2.71)             & 51.64  ($\pm$3.53)             \\
			DABR(CTV)                                & 94.14 ($\pm$0.67)             & 82.66 ($\pm$1.78)             & 75.06($\pm$1.67)              & 59.40 ($\pm$1.93)             & 91.67 ($\pm$1.05)             & 81.34 ($\pm$2.80)             & 73.69 ($\pm$2.39)             & 56.38  ($\pm$2.24)             \\
			DABR(MTA)                                & 93.64 ($\pm$0.93)             & 83.92 ($\pm$1.58)             & 73.42($\pm$2.28)              & 61.29 ($\pm$2.09)             & 90.82 ($\pm$1.37)             & 80.95 ($\pm$2.59)             & 71.77 ($\pm$2.80)             & 53.29  ($\pm$3.33)             \\
			\midrule
			DABR(MDV)+CQSBE                           & 96.18 ($\pm$1.21)             & 88.58 ($\pm$2.64)             & 79.08($\pm$4.87)              & 63.03 ($\pm$2.05)             & 91.89 ($\pm$1.43)             & 84.87 ($\pm$4.42)             & 70.76 ($\pm$2.38)             & 55.22  ($\pm$3.30)             \\
			DABR(CTV)+CQSBE                           & 96.75 ($\pm$0.61)             & 89.81 ($\pm$2.81)             & 81.56($\pm$4.41)              & 64.56 ($\pm$4.78)             & 96.15 ($\pm$0.68)             & 84.60 ($\pm$2.75)             & 73.85 ($\pm$2.09)             & 63.76  ($\pm$2.55)             \\
			DABR(MTA)+CQSBE                           & 97.16 ($\pm$0.81)             & 88.74 ($\pm$2.04)             & 83.10($\pm$3.07)              & 64.37 ($\pm$4.67)             & 94.81 ($\pm$0.98)             & 84.47 ($\pm$3.46)             & 72.74 ($\pm$1.89)             & 63.18  ($\pm$2.51)             \\
			\midrule
			DABR(MDV)+CQSBE+nBAS                      & \textbf{100} ($\pm$0.00)      & \underline{99.98} ($\pm$0.04) & \underline{99.95} ($\pm$0.01) & 98.37 ($\pm$0.08)             & 99.80 ($\pm$0.20)             & 99.72 ($\pm$0.26)             & 98.36 ($\pm$0.54)             & \underline{93.72}  ($\pm$1.60) \\
			DABR(CTV)+CQSBE+nBAS                      & \textbf{100} ($\pm$0.00)      & \textbf{100} ($\pm$0.00)      & 99.94 ($\pm$0.01)             & \textbf{98.72} ($\pm$0.10)    & \textbf{99.90} ($\pm$0.15)    & \textbf{99.76} ($\pm$0.20)    & \textbf{98.68} ($\pm$0.40)    & \textbf{95.25} ($\pm$1.37)     \\
			DABR(MTA)+CQSBE+nBAS                      & \textbf{100} ($\pm$0.00)      & \textbf{100} ($\pm$0.00)      & \textbf{99.96} ($\pm$0.01)    & \underline{98.68} ($\pm$0.09) & \underline{99.89} ($\pm$0.13) & \underline{99.73 }($\pm$0.24) & \underline{98.50} ($\pm$0.40) & 93.18  ($\pm$1.29)             \\
			\toprule[2pt]
		\end{tabular}
	}
	\raggedright \\
	\footnotesize \textbf{Bold} numbers indicate optimal results, while \underline{underlined} numbers denote sub-optimal results.
	\vspace{-1em}
\end{table*}

\subsection{Resiliency}
\label{exp:res}

\tab\ref{tab:reach} summarizes the reachability performance of different routing schemes under varying ISL failure rates. 
The baseline methods degrade significantly under high failures, dropping to 44–49\% reachability at 30\% failure ratio. 
In contrast, BlockFlex demonstrates progressively stronger resilience as its components are integrated: using only the basic DABR improves reachability to 51–61\%; adding CQS-based block evolution further increases it to 55–64\%; and with the full system including nBAS protection, reachability reaches 93–98\% under the same severe conditions. 
This superior performance stems from BlockFlex’s grouping of satellites into blocks, which provides richer connectivity and reduces susceptibility to local minima.
When routing does encounter dead ends, the integrated nBAS mechanism enables efficient backtracking and path recovery. 
These features allow the complete BlockFlex system (DABNet+CQSBE+nBAS) to achieve 93–98\% reachability even under 30\% ISL failures—nearly doubling the performance of baseline schemes and demonstrating robust resilience in highly dynamic MCN environments.

Additionally, the baseline schemes, even in non-failure scenarios, suffer from routing unreachability due to local minima \cite{henderson2000distributed}. 
When a packet reaches a satellite that is locally optimal but still distant from the destination, forwarding fails.
In contrast, BlockFlex eliminates such unreachability in Starlink under non-failure conditions. If the destination lies within the same block as the current forwarding unit, intra‑block routing delivers the packet directly. If the destination is outside the block, inter‑block geographic routing operates over a larger decision scope, effectively bypassing local minima and ensuring end‑to‑end reachability.

\subsection{Efficiency}
\label{exp:eff}

\subsubsection{Routing Rediscovery}

\begin{figure}[t!]
    \begin{center}
        \subfloat[Num. of routing re-discovery in various routing schemes.]
        {
            \includegraphics[width=1\linewidth]{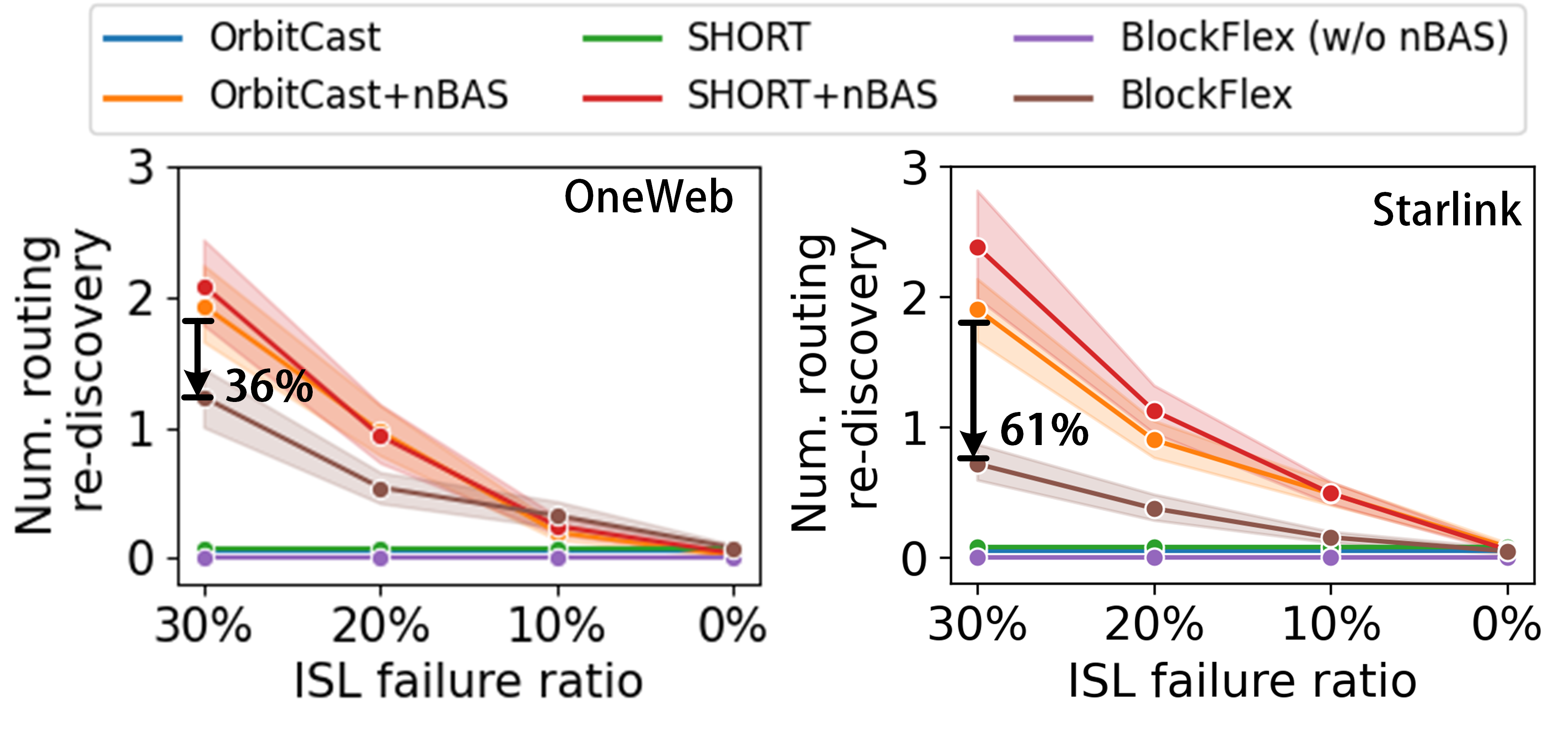}
        }
        
        \subfloat[FIB updates.]
    {
        \includegraphics[width=0.47\linewidth]{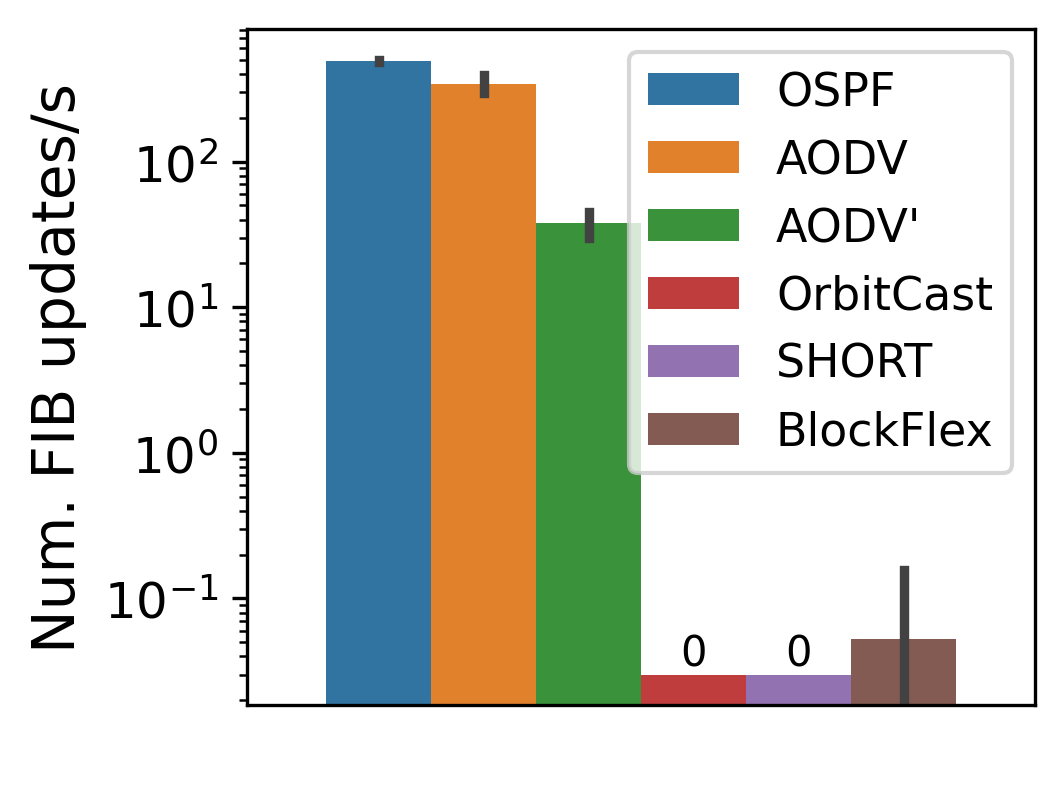}
    }\subfloat[Control message generation.]
    {
    \includegraphics[width=0.47\linewidth]{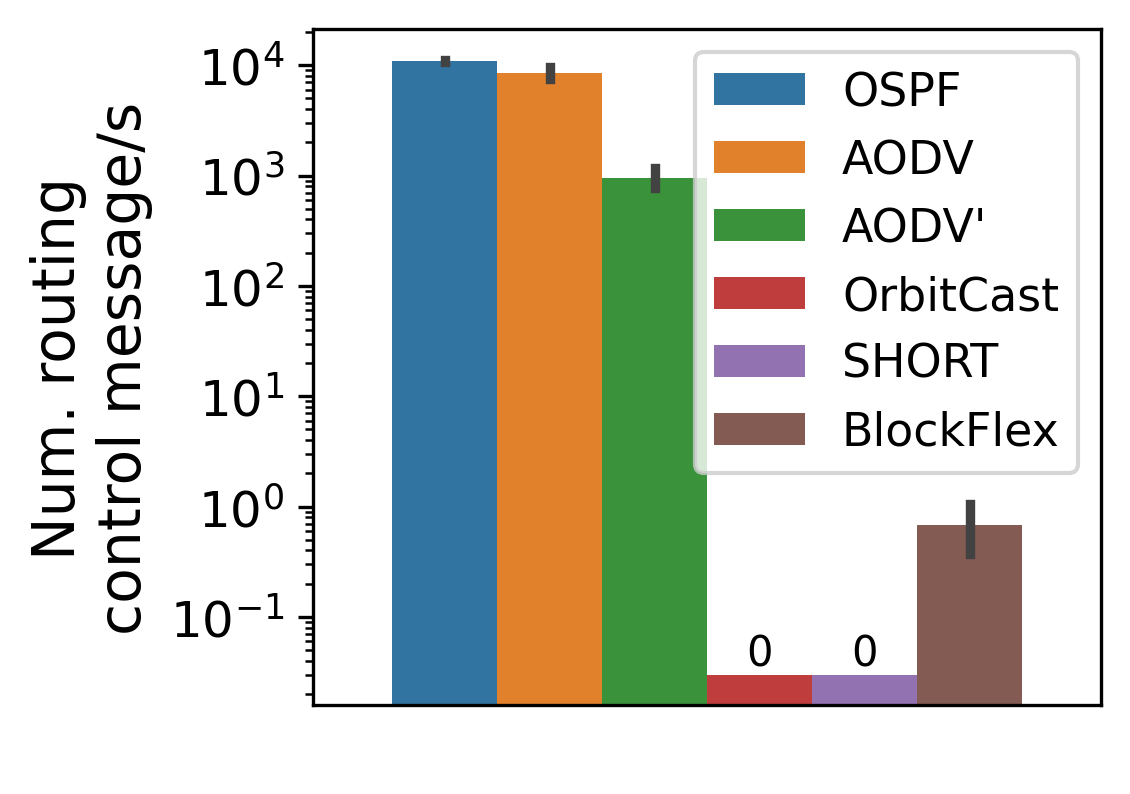}
    }
\end{center}
    \caption{Routing convergence overheads.} 
\vspace{-1em}
       \label{fig:eff}
   \end{figure}

\fig\ref{fig:eff} (a) compares the number of routing rediscovery events per dataflow across different routing schemes.
While the nBAS protection mechanism substantially improves routing reachability, it also introduces additional rediscovery procedures. 
The baseline OrbitCast, SHORT, and BlockFlex (w/o nBAS) schemes exhibit no rediscovery, maintaining a count of zero. In contrast, enabling nBAS triggers protection routing through the cached PFS upon packet delivery failure, which activates rediscovery operations. 
In our experiments, the maximum number of backward acknowledgment attempts per route was uniformly set as $n_{\text{max}}=5$.
Compared to the nBAS enhanced versions of OrbitCast and SHORT, BlockFlex's aggregation of forwarding satellites into blocks reduces the number of fallback signaling processes during protection forwarding. 
Compared to the suboptimal method (OrbitCast), the proposed BlockFlex scheme achieves a reduction in average routing rediscovery count of 36\% (1.23 vs. 1.93) on OneWeb and 61\% (0.32 vs. 0.84) on Starlink, thereby lowering overall routing overhead.

\subsubsection{Routing Convergence Overhead}
In \fig\ref{fig:eff} (b) and (c), routing convergence overhead is quantified by counting topology changes and the number of traffic flows within the convergence scope, while considering protocol-specific characteristics to compute the frequency of FIB updates and the number of control messages generated during convergence. All routing schemes in the simulation experience the same traffic load, except AODV, which operates at 50\% of the load of other schemes.
Due to satellite-ground dynamics and intermittent ISL failures, OSPF triggers frequent network-wide FIB updates and floods the system with excessive control messages. AODV, with its on-demand routing nature, incurs significantly lower overhead than OSPF at 50\% load—approximately 10\% of OSPF’s overhead (green bar). However, its overhead approaches that of OSPF as traffic load increases to the same level.
OrbitCast~\cite{lai2021orbitcast} and SHORT~\cite{li2024stable} leverage stateless forwarding and convergence-free routing, resulting in zero FIB updates and control message generation.
The overall overhead of BlockFlex stems entirely from the aggregated re-convergence costs of all evolving blocks within the network. 
It confines convergence to localized blocks, maintaining a 0.2\% FIB update frequency and 0.1\% control message overhead relative to OSPF.
While convergence-free schemes achieve zero overhead, BlockFlex significantly outperforms convergent protocols such as OSPF and AODV, despite the resilience limitations of geographic routing as shown in \Sec\ref{exp:res}.

\subsubsection{Routing Computation Time Consumption}

\begin{figure}[t!] 
    \begin{center}
        \subfloat[Time consumption fluctuation.]
        {
                \includegraphics[width=0.5\linewidth]{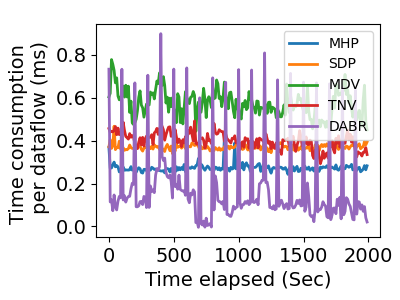}
        }\subfloat[Time consumption distribution.]
        {
                \includegraphics[width=0.5\linewidth]{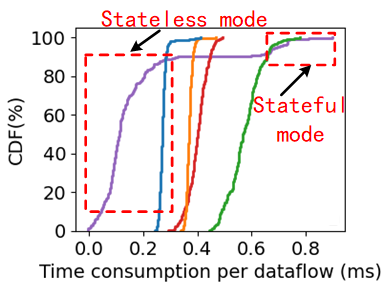} 
        }
        \end{center}

        \begin{center}

        \subfloat[Avg. time consumption (ms).]
        {
\renewcommand{\arraystretch}{1.1}  
             \scalebox{0.8}{
                \begin{tabular}{lccccc}
                    \toprule[1pt]
                      & MHP & SDP & SHORT & OrbitCast & BlockFLEX \\
                    \midrule
                    Time (ms) & 0.272 & 0.369 & 0.572 & 0.399 & \textbf{0.174} \\
                    Std (ms) & $\pm$0.017 & $\pm$0.012 & $\pm$0.036 & $\pm$0.061 & $\pm$0.189 \\
                    \bottomrule[1pt]
                    \end{tabular}
            }
        }
         \end{center}

    \caption{Comparison of time consumption under different routing schemes
    } 
   \label{fig:time}
\end{figure}

\udm{The time consumption of routing computation is analyzed under a fixed dataflow configuration}, with the PFS update period in BlockFlex set to $\delta = 100~\text{s}$. Only the computational overhead during the path selection phase is measured. Both the Minimum Hop-count Path (MHP) and Shortest Distance Path (SDP) are computed using the \texttt{networkx} library \cite{hagberg2008exploring}. These two path selection methods are widely adopted in protocols such as AODV \cite{perkins2003rfc3561} and OSPF \cite{rfc5340}.
As illustrated in \fig\ref{fig:time}, despite operating on networks with thousands of nodes, both MHP and SDP maintain exceptionally low computational overhead via the \texttt{networkx} library. Within geographic routing schemes, SHORT requires more time than OrbitCast due to its denser vector operations.

In BlockFlex, \fig\ref{fig:time} (a) reveals a periodic pattern in computation time, with cycles corresponding to the predefined PFS recomputation period ($\delta = 100~\text{s}$). 
\fig\ref{fig:time} (b) demonstrates that the stateful mode maintains low computation time, as next-hop decisions are efficiently determined through simple table lookups. In contrast, the stateless mode incurs substantially higher computational overhead, as each satellite along the traffic path must perform computationally intensive geographic-based PFS calculations.
In the experiments, the SPQ computation is triggered by newly adjacent source satellite sets. Its time cost is negligible compared to PFS recomputation, resulting in less noticeable periodicity. 
\udm{As clearly demonstrated in \fig\ref{fig:time} (c), BlockFlex achieves significant computational efficiency: it reduces average computation time by 56\% compared to the geographic routing scheme OrbitCast ($0.174$~ms vs. $0.399$~ms), and by 36\% compared to the shortest-path routing MHP ($0.174$~ms vs. $0.274$~ms).}
Additionally, it is anticipated that as the update interval $\delta$ in BlockFlex increases, the overall computational overhead will decrease due to reduced spatial computations; however, this will result in degraded routing reachability.
\subsection{Latency and Stability}
\label{exp:lat}

\subsubsection{Latency Analysis in Global View}
\begin{figure}[t!]
    \begin{center}
    \subfloat[Path stretch (normalized latency) of traffics.]
    {
        \begin{minipage}{0.5\textwidth}
        \includegraphics[width=\linewidth]{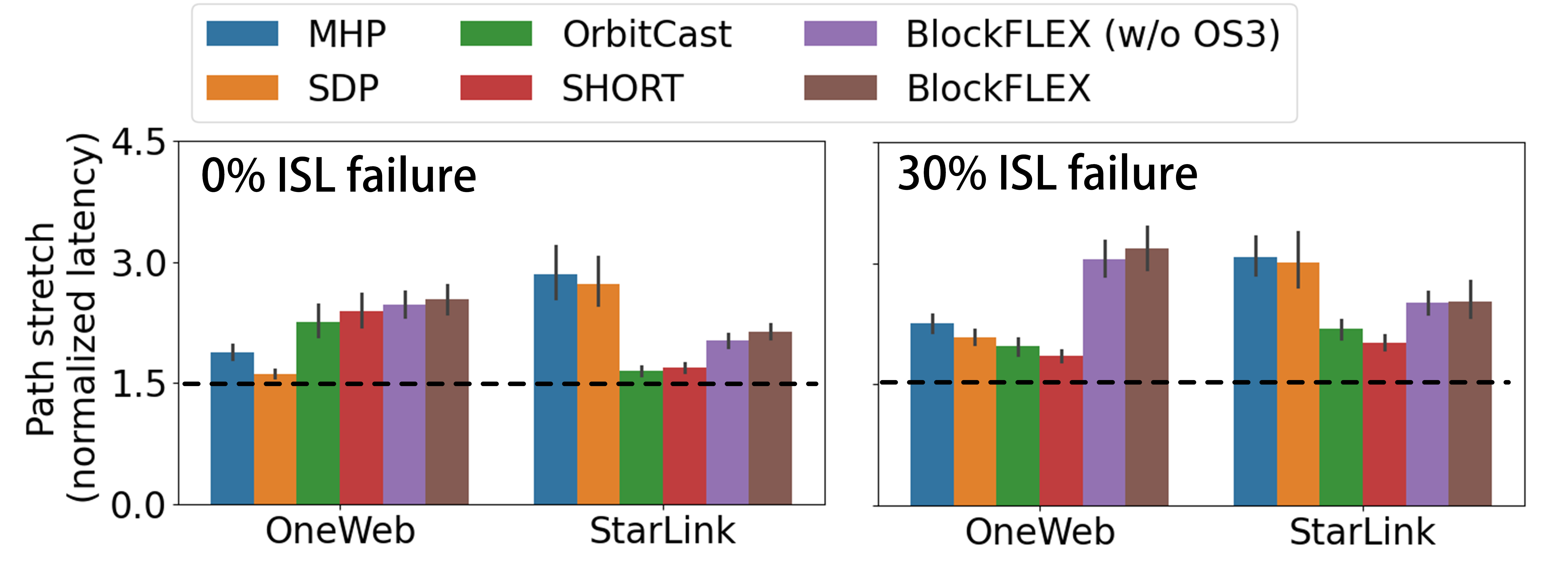}
        \end{minipage}
    }

    \vspace{-1em}
    \subfloat[Latency jitter of traffics.]
    {
    \begin{minipage}{0.5\textwidth}
    \includegraphics[width=\linewidth]{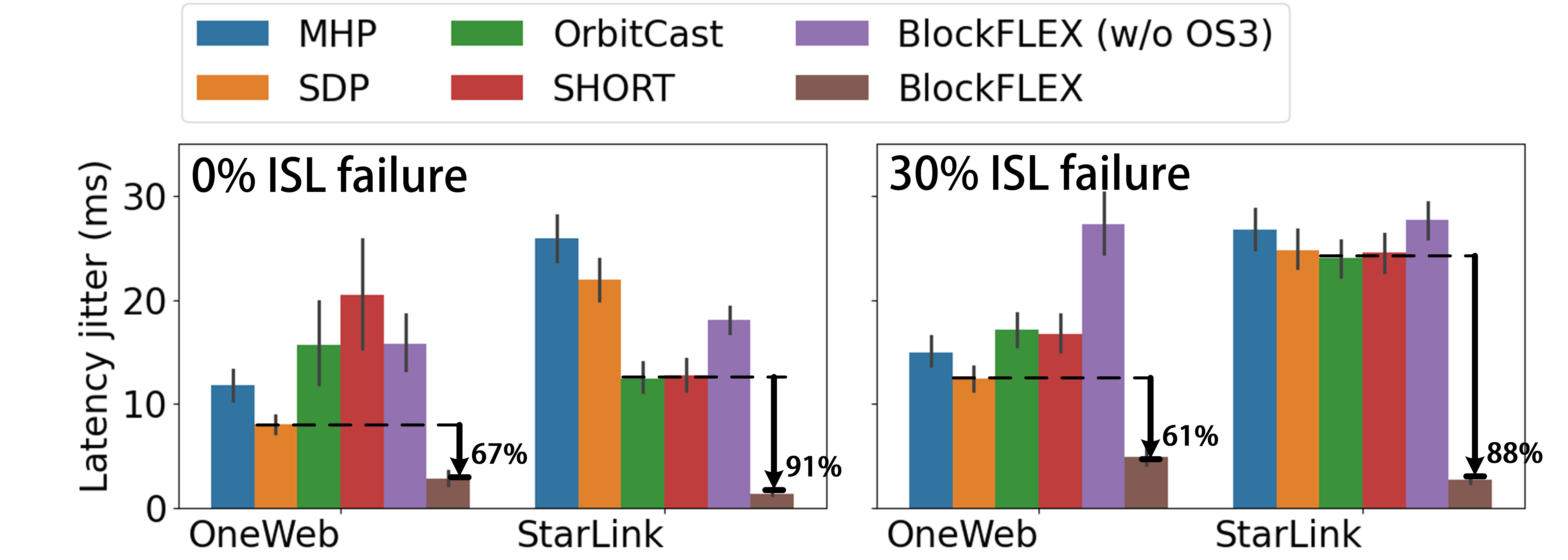}
    \end{minipage}
    }

\end{center}
\vspace{-1em}
    \caption{Latency and path stability.} 
\vspace{-1em}
       \label{fig:route_stab}
   \end{figure}

\fig\ref{fig:route_stab} compares the one-way propagation latency across different routing schemes.
Path stretch, defined as the ratio between the actual path length and the geographic distance, serves as a normalized latency metric \cite{bhattacherjee2019network}.
Since the speed of light in optical fiber is approximately $2c/3$, where $c$ represents the speed of light in vacuum, any stretch value below 1.5 implies performance exceeding that of terrestrial direct fiber links
(as indicated by the dashed line in Fig.~\ref{fig:route_stab}(a)).

 Compared to shortest-path routing (SDP, MHP) and geographic routing (SHORT, OrbitCast), BlockFlex does not exhibit significant advantages in latency, yet it achieves comparable results.
This can be attributed to two main factors. First, BlockFlex considers a larger set of satellites during path selection, enhancing path diversity but also increasing the likelihood of higher-latency paths.
Second, conventional average latency metrics often exclude routing failures, which occur frequently in MCNs as discussed in \Sec\ref{exp:res}, potentially leading to underestimated latency values.
Additionally, after incorporating the \texttt{OS3} mechanism, BlockFlex does not exhibit a significant advantage in latency. This is attributed to the fact that while \texttt{OS3} stabilizes path diversity, it also excludes certain source satellites that, despite having shorter available time windows, could provide shorter paths.
Nevertheless, BlockFlex consistently demonstrates lower latency jitter, as shown in Fig. \ref{fig:route_stab} (b). It reduces jitter by at least 67\% compared to the second‑best scheme in scenarios with no ISL failures, and maintains a reduction of at least 61\% even under 30\% ISL failure conditions.


\subsubsection{Latency Analysis in Local View}

\begin{figure}[t!]
    \begin{center}
    \subfloat[End-to-end one-way latency between various cities.]
    {
        \begin{minipage}{0.5\textwidth}
        \includegraphics[width=\linewidth]{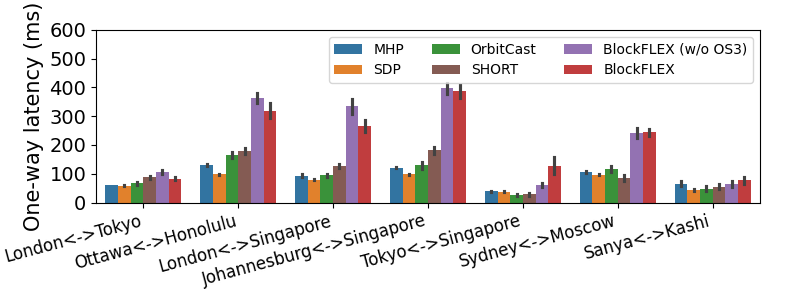}
        \end{minipage}
    }

    \subfloat[Latency fluctuation between London and Singapore.]
    {
    \begin{minipage}{0.5\textwidth}
    \includegraphics[width=\linewidth]{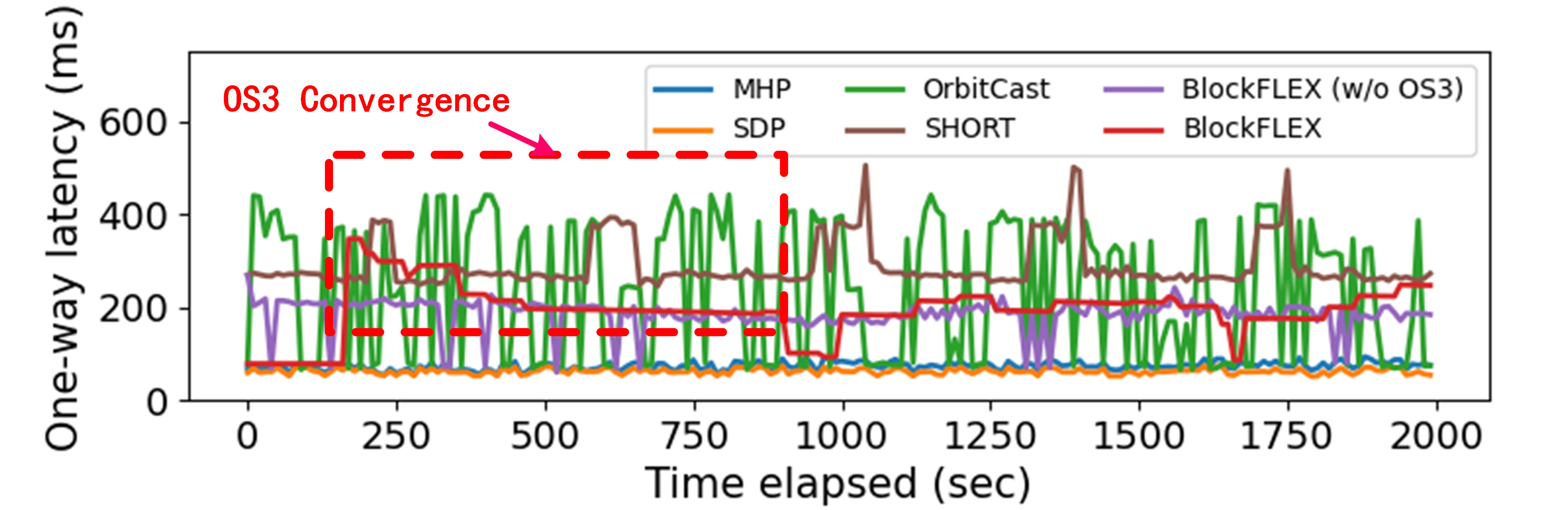}
    \end{minipage}
    }
\end{center}
    \caption{Latency in end-to-end session.} 
       \label{fig:local_route}
   \end{figure}

\fig\ref{fig:local_route} (a) shows selected end‑to‑end data flows in the OneWeb constellation.
The latencies of flows such as Ottawa-Honolulu, London–Singapore, and Johannesburg–Singapore are higher under BlockFlex. This illustrates that although the OS3 mechanism reduces latency and jitter overall, it can introduce a noticeable convergence phase, as indicated by the red dash box in \fig\ref{fig:local_route} (b). This temporary convergence raises the average latency for some flows, though it gradually decreases to a reasonable level once convergence completes.
On the other hand, BlockFlex exhibits significantly more stable latency across flows. Consequently, MCNs employing BlockFlex can effectively mitigate packet reordering, thereby substantially improving the Quality of Service for latency‑sensitive applications such as streaming services.

\section{Conclusion}
\label{sec:con}
In this paper, BlockFlex is presented — a new routing architecture that integrates a robust overlay (DABNet) and a hybrid routing scheme (DABR) to improve both resilience and efficiency for dynamic mega-constellation networks.
The evaluation shows that BlockFlex maintains strong reachability even under severe ISL failures, while keeping routing overhead very low compared to existing state‑of‑the‑art approaches.
Looking ahead, extending this design to multi‑shell MCNs will require adapting DABNet's evolution and DABR's routing to inter‑shell dynamics. Further work in this direction will be important for future scalable and adaptive satellite networking.

\section{Data Availability}
The source code and data have been anonymized and are available at \url{https://github.com/wxton123/BlockFlex}
\bibliographystyle{IEEEtran}
\bibliography{ref.bib}

\end{document}